\input{amssymb.sty}
\documentclass[aps,twocolumn,pre,showpacs,amsmath,amssymb]{revtex4}
\usepackage{graphicx}
\usepackage{dcolumn}
\usepackage{bm}
\usepackage{epsfig}

\def\kb{k_{\mbox{\scriptsize B}}}

\def\rhodiam{\rho_{\mbox{\scriptsize diam}}}

\begin{document}
\bibliographystyle{prsty}

\title{Yang-Yang Anomalies and Coexistence Diameters: Simulation of Asymmetric Fluids}

\author{Young C. Kim}
\affiliation{Institute for Physical Science and Technology, University of Maryland,
College Park, Maryland 20742}

\date{\today}

\begin{abstract}
A general method for estimating the Yang-Yang ratio, ${\cal R}_{\mu}$, of a model fluid via Monte Carlo simulations is presented on the basis of data for a hard-core square-well (HCSW) fluid and the restricted primitive model (RPM) electrolyte. The isothermal minima of $Q_{L}\equiv\langle m^{2}\rangle^{2}_{L}/\langle m^{4}\rangle_{L}$ are evaluated at $T_{c}$ in an $L\times L\times L$ box where $m = \rho - \langle\rho\rangle_{L}$ is the density fluctuation. The ``complete'' finite-size scaling theory for the $Q_{\mbox{\scriptsize min}}^{\pm}(T_{c};L)$ incorporates pressure mixing in the scaling fields, thereby allowing for a Yang-Yang anomaly. It yields a dominant term in the asymmetry, $Q_{\mbox{\scriptsize min}}^{+}-Q_{\mbox{\scriptsize min}}^{-}$, varying as $L^{-\beta/\nu}$ with an amplitude proportional to the  crucial pressure-mixing coefficient, $j_{2}$. The reliably known critical order-parameter distribution for $(d=3)$ Ising systems then enables one to estimate $j_{2}$, thereby yielding ${\cal R}_{\mu}$, from the $Q$-minima together with information on the nonuniversal amplitudes for the order-parameter and the susceptibility. The detailed analysis needed to estimate $j_{2}$ for an HCSW fluid and the RPM is presented. Furthermore, the $Q$-minima below $T_{c}$ can also provide the {\em coexistence-curve diameters}, $\rho_{\mbox{\scriptsize diam}}(T)\equiv \frac{1}{2}(\rho^{+}+\rho^{-})$, very close to $T_{c}$ for both models: here $\rho^{\pm}(T)$ are the densities of the coexisting liquid and gas phases. The recently developed recursive scaling algorithm for $\Delta\rho_\infty (T)\equiv \rho^{+}-\rho^{-}$ is adapted to investigate the corresponding universal scaling functions. The two extremal forms of these scaling functions are computed with the aid of the exactly soluble decorated lattice-gas model. The critical densities for the RPM and HCSW fluid found via this route are consistent with previous estimates obtained from the data above $T_{c}$; the magnitudes of the leading $|T-T_c|^{2\beta}$ and $|T-T_c|^{1-\alpha}$ corrections to $\rho_{\mbox{\scriptsize diam}}(T)$ are estimated.

\end{abstract}
\maketitle

\section{\label{sec1}Introduction}

The presence of a Yang-Yang anomaly means that the chemical potential, $\mu_{\sigma}(T)$, on a phase boundary separating liquid and vapor exhibits a divergence of the second temperature derivative, $d^{2}\mu_{\sigma}(T)/dT^{2}$, when the critical temperature $T_c$ of the fluid is approached from below in the same fashion as the constant-volume specific heat, $C_{V}(T;\rho_{c})$. This possibility was first proposed by Yang and Yang \cite{yan:yan} 40 years ago on the basis of a simple thermodynamic relation they derived (referred to later as the Yang-Yang equation) which connects the specific heat to the pressure and chemical potential derivatives. It was only more recently, however, that the Yang-Yang anomaly in fluid criticality was seriously investigated by Fisher and coworkers \cite{fis:ork,ork:fis:ust}. They analyzed carefully  experimental data for the two-phase heat-capacity of propane (C$_3$H$_8$) and CO$_2$ in the critical region and showed that $d^2 \mu_{\sigma}/dT^2$ {\em indeed} diverges like the specific heat at criticality. Further study of the experimental data for propane showed that impurities in the system can have significant effects on the heat-capacity data \cite{kos}; however, the existence of Yang-Yang anomalies was not ruled out. Nonetheless, further careful, experimental investigations are desirable \cite{ani:zho:bar}.

Recently computer simulations have become an efficient tool to study the behavior of fluids and complex fluids and, in particular, to enhance our understanding of their critical behavior. Investigating a Yang-Yang anomaly in fluid criticality can thus be aided by simulations for model fluids, since impurities are absent in such models; but such simulations pose a serious challenge. Specifically, any sharp divergence of a thermodynamic quantity will be rounded owing to finite-size effects. These arise from the divergence of the correlation length at criticality. In fact, Orkoulas, Fisher and Panagiotopoulos \cite{ork:fis:pan} performed grand canonical Monte Carlo simulations on a hard-core square-well (HCSW) model fluid and concluded that this model probably exhibits a negative but small Yang-Yang anomaly quantified by a Yang-Yang ratio (see below) ${\cal R}_{\mu}=-0.08\pm 0.12$; however, they could not rule out the possibility that a Yang-Yang anomaly was absent. Thus, one might ask: How might one measure a Yang-Yang anomaly precisely from simulations on model fluids? 

To account for a Yang-Yang anomaly, it is necessary to mix the pressure deviation, $p-p_c$, into the asymptotic scaling fields. More specifically, the full thermodynamics of a one-component fluid near the bulk critical point $(p_{c},T_{c},\mu_{c})$ can be described with three relevant scaling fields \cite{fis:ork,kim:fis:ork}, namely,
  \begin{eqnarray}
    \tilde{p} & = & \check{p} - k_{0}t - l_{0}\check{\mu} + \cdots,  \label{eq1.1} \\
    \tilde{t} & = & t - l_{1}\check{\mu} - j_{1}\check{p} + \cdots,  \label{eq1.2} \\
    \tilde{h} & = & \check{\mu} - k_{1}t - j_{2}\check{p} + \cdots,  \label{eq1.3}
  \end{eqnarray}
where the dimensionless deviations of the thermodynamic fields from their critical values are
  \begin{equation}
   t \equiv \frac{ T-T_{c}}{T_{c}}, \hspace{0.2in} \check{p} \equiv \frac{p-p_{c}}{\rho_c\kb T_c}, \hspace{0.2in} \check{\mu}\equiv \frac{\mu-\mu_c}{\kb T_c},  \label{eq1.4}
  \end{equation}
while $j_{1},j_{2},k_{0},\cdots$ are nonuniversal mixing coefficients and $\rho_c$ is the critical density. The scaling hypothesis then asserts \cite{fis:ork,kim:fis:ork}
  \begin{equation}
   \tilde{p}\approx Q|\tilde{t}|^{2-\alpha} W_{\pm}(U\tilde{h}/|\tilde{t}|^{\Delta}),\hspace{0.3in} \mbox{for $\tilde{t}\gtrless 0$},  \label{eq1.5}
  \end{equation}
where $\alpha$ is the critical exponent for the specific heat while $\Delta$ is the gap exponent satisfying the relation $\Delta = \beta + \gamma$. Here $\beta$ and $\gamma$ are the critical exponents for the order parameter and the susceptibility/compressibility, respectively. In this formulation $W_{\pm}(x)$ represents two branches of a universal scaling function while $Q$ and $U$ are nonuniversal amplitudes depending on microscopic details of the system under study. Note, however, that we have neglected here, for simplicity, both corrections to scaling and higher order nonlinear mixing terms \cite{kim,kim:fis:ork}.

The strength of a Yang-Yang anomaly is conveniently measured by ${\cal R}_{\mu}$, the limiting ratio of $\tilde{C}_{\mu}(T)\equiv -T(d^{2}\mu_{\sigma}/dT^{2})$ to the constant-volume specific heat, $C_{V}(T,\rho=\rho_{c})$ (where $\rho$ is the number density). It then follows that the Yang-Yang ratio ${\cal R}_{\mu}$ is related \cite{kim:fis:ork,kim} to the pressure-mixing coefficient $j_{2}$ via
 \begin{equation}
  {\cal R}_{\mu}=-j_{2}/(1-j_{2}).  \label{eq.Rmu}
 \end{equation}

To measure the Yang-Yang ratio, ${\cal R}_{\mu}$, quantitatively via simulations, the first question is: What thermodynamic or other quantities should be studied? The answer is not obvious since a direct attempt \cite{ork:fis:pan} to study $\tilde{C}_{\mu}(T)$, etc.\ proves not effective. Here we show that the desired information can be obtained by carefully investigating the finite-size-system parameter $Q_{L}$ defined by \cite{bin,blo:lui:her,kim:fis:lui}
  \begin{equation}
    Q_{L}(T;\langle\rho\rangle_{L}) \equiv \langle m^{2}\rangle_{L}^{2}/\langle m^{4}\rangle_{L}, \hspace{0.2in} m = \rho - \langle\rho\rangle_{L},  \label{eq1.6}
  \end{equation}
where $\langle \cdot \rangle_{L}$ is the grand canonical average at fixed $T$ and $\mu$ chosen to yield the desired mean density. In the thermodynamic limit $(L\rightarrow\infty)$, the parameter $Q_{L}$ then exhibits surprising, singular behavior on the two sides of the coexistence curve \cite{rov:hee:bin,dun:lan,kim:fis}; namely it {\em vanishes} on the coexistence curve boundary $\rho=\rho^{+}(T)$ and $\rho=\rho^{-}(T)$, where $\rho^{\pm}(T)$ are the densities of the coexisting liquid and gas states in the two-phase region. However, in the one-phase region, $Q_\infty$ remains $\frac{1}{3}$ but drops discontinuously to zero at the phase boundary when the two-phase region is entered; it then rises continuously to unity as the mean density approaches the coexistence-curve diameter \cite{kim:fis}, namely,
 \begin{equation}
  \rho_{\mbox{\scriptsize diam}}(T) \equiv \mbox{$\frac{1}{2}$}(\rho^{+}+\rho^{-}). \label{eq.diam}
 \end{equation}

Of course in a finite system, these discontinuities become rounded. Thus $Q_{L}(\langle\rho\rangle_L)$ exhibits two isothermal minima of magnitudes, say $Q_{\mbox{\scriptsize min}}^{\pm}(T;L)$, at densities, $\rho_{\mbox{\scriptsize min}}^{\pm}(T;L)$, near the coexistence curve boundary \cite{kim:fis,bin:lan}. (See also Fig.\ \ref{fig1} below.)

For symmetric systems such as Ising models or lattice gas models, the two minima, $Q_{\mbox{\scriptsize min}}^{\pm}(T;L)$, have equal height while their corresponding densities, $\rho_{\mbox{\scriptsize min}}^{\pm}(T;L)$, are located symmetrically about the critical density, $\rho=\rho_{c}$. However, as soon as mixing enters in the scaling fields $\tilde{t}$ and $\tilde{h}$ [see (\ref{eq1.2}) and (\ref{eq1.3})], one finds that $Q_{L}(\langle\rho\rangle_L)$ becomes asymmetric. In fact, according to the {\em complete} finite-size scaling theory (an extension of (\ref{eq1.5}) to finite systems \cite{kim:fis}), the minima, $Q_{\mbox{\scriptsize min}}^{\pm}$, and their locations, $\rho_{\mbox{\scriptsize min}}^{\pm}$, exhibit leading asymmetric contributions arising from the pressure-mixing coefficient $j_{2}$. More specifically, let us define normalized {\em asymmetry factors} of the minima via
  \begin{eqnarray}
    {\cal A}_{\mbox{\scriptsize min}}(T;L) & \equiv & (Q_{\mbox{\scriptsize min}}^{+}-Q_{\mbox{\scriptsize min}}^{-})/(Q_{\mbox{\scriptsize min}}^{+}+Q_{\mbox{\scriptsize min}}^{-}),  \label{eq1.7}\\
    {\cal B}_{\mbox{\scriptsize min}}(T;L) & \equiv & \frac{\rho_{\mbox{\scriptsize min}}^{+}+\rho_{\mbox{\scriptsize min}}^{-}-2\rho_{\mbox{\scriptsize diam}}}{\rho_{\mbox{\scriptsize min}}^{+}-\rho_{\mbox{\scriptsize min}}^{-}}. \label{eq1.8}
  \end{eqnarray}
It is obvious that these quantities vanish identically for symmetric systems. For asymmetric situations, however, both exhibit leading finite-size correction terms varying as $L^{-\beta/\nu}$ which are proportional to $j_{2}$, while a nonvanishing $(l_{1}+j_{1})$ combination induces a further correction term decaying as $L^{-(\Delta -1)/\nu}$: see below in Sec.\ \ref{sec3}. Here $\nu$ is the correlation-length exponent. Note that for the $(d=3)$-dimensional Ising universality class to which fluid criticality is believed to belong, we have $\beta/\nu\simeq 0.517$ and $(\Delta-1)/\nu\simeq 0.897$: see Table \ref{tab1}. 
\begin{table}[ht]
\vspace{-0.1in}
\caption{\label{tab1}Preferred values for the universal Ising $d=3$ critical exponents adopted here \cite{fis:zin}.}
\vspace{0.1in}
\begin{ruledtabular}
\begin{tabular}{ccccccc}
  $\alpha$ & $\beta$ & $\gamma$ & $\delta$ & $\Delta$ & $\nu$ & $\theta$\footnotemark[1] \\ \hline
  0.109 & 0.326 & 1.239 & 4.80 & 1.565 & 0.630 & 0.52 \\
\end{tabular}
\end{ruledtabular}
\footnotetext[1]{From Ref.\ \cite{gui:zin}.}
\end{table}
It will be shown in Sec.\ \ref{sec3} that the amplitudes of these leading terms can be expressed in terms of the nonuniversal order-parameter and susceptibility amplitudes (which, in principle, can be measured from simulations) and certain {\em universal} constants. This then raises the hope of measuring $j_{2}$ (or ${\cal R}_{\mu}$) via simulations.

Indeed, it is shown below that the critical order-parameter distribution function for Ising systems and the estimated nonuniversal amplitudes, $B$ and $C^{+}$, for the density discontinuity and the susceptibility, provide precise estimates for the pressure-mixing coefficient, $j_{2}$, and, thereby via (\ref{eq.Rmu}) for the Yang-Yang ratio, ${\cal R}_{\mu}$. One should notice, however, that measuring $j_{2}$ requires precise knowledge of $T_{c}$ and $\rho_c$ which must also be found by well designed simulations utilizing finite-size scaling theory \cite{kim:fis:lui,kim:fis}.

Among the consequences of pressure mixing in the scaling fields (beyond the divergence of $d^{2}\mu_{\sigma}/dT^{2}$ at criticality) are the appearance of a complex spectrum of singular correction exponents in various thermodynamic properties \cite{kim,kim:fis:ork}. In particular, the pressure-mixing coefficient $j_{2}$ induces a leading singular term varying as $|t|^{2\beta}$ in the coexistence-curve diameter, $\rho_{\mbox{\scriptsize diam}}(T)$ \cite{kim:fis:ork}. This term then dominates the previously known $|t|^{1-\alpha}$ term \cite{mer}. Hence the nonvanishing of $j_{2}$ may affect the shape of the diameter strongly near criticality, and this inevitably hampers the precise estimation of the critical density.

A conventional way of estimating the coexisting liquid and gas densities, $\rho^{\pm}(T)$, is to observe the grand canonical equilibrium distribution function, ${\cal P}_{L}(\rho)$, of the density, $\rho$, below the critical temperature, $T_c$. In a finite geometry of dimensions $L^d$ with periodic boundary conditions the distribution function generally exhibits two peaks near $\rho^{\pm}(T)$ when $L\gg a$ where $a$ measures the size of particles \cite{bin,bin:lan,bor:kap}. For sufficiently large $L$, these peaks can be represented by two Gaussians clearly distinguished from each other. Observing these peaks then provides direct estimates of $\rho^{\pm}(T)$ when $T$ is sufficiently low; for fluids which exhibit no ``obvious symmetry,'' the equal-weight prescription has been widely accepted to estimate $\rho^{\pm}(T)$ from ${\cal P}_{L}(\rho)$ \cite{bin:lan,bor:kap}. This approach has yielded reasonable estimates of $\rho^{\pm}(T)$ for the HCSW fluid \cite{ork:fis:pan} and for the RPM electrolyte \cite{lui} but only up to $1$-$2\%$ below $T_c$. However, when $T$ approaches $T_c$ more closely, finite-size effects blur the distinction between the coexisting liquid and gas states for most computationally accessible system sizes thereby seriously hampering the reliable estimation of the coexistence curve. Furthermore, ${\cal P}_{L}(\rho)$ at criticality for finite $L$ still exhibits two well-separated maxima; on the other hand, $\rho^{\pm}(T)$ should (for $\beta >0$) merge precisely at the critical density, $\rho_c$. Hence, unless one can simulate sufficiently large system sizes which exceed the correlation length, it is almost impossible to obtain reliable estimates of $\rho^{\pm}(T)$ in the critical region via this route; to do so would require a rather full understanding of finite-size effects on the distribution function.

To meet this challenge, a scaling algorithm has been developed recently for estimating the liquid and gas coexisting densities, $\rho^{\pm}(T)$, of model fluids from grand canonical simulation data \cite{kim:fis:lui,kim:fis2}. The algorithm utilizes data $\{Q_{\mbox{\scriptsize min}}^{\pm}(T;L),\rho_{\mbox{\scriptsize min}}^{\pm}(T;L)\}$ for the $Q$-minima and derives the density discontinuity, 
 \begin{equation}
  \Delta\rho_{\infty}(T)\equiv (\rho^{+}-\rho^{-}), \label{eq.discon}
 \end{equation}
by constructing an appropriate universal finite-size scaling function. By this route precise estimates for $\Delta\rho_{\infty}(T)$ for the HCSW fluid and the RPM electrolyte were obtained up to temperatures only $0.01$-$0.1\%$ below $T_c$ \cite{kim:fis:lui,kim:fis2}. On the other hand, estimating the diameter, $\rho_{\mbox{\scriptsize diam}}(T)$, turns out to be more complicated as will be demonstrated in the second part of this article.

In order to calculate the coexistence-curve diameter, we will compare the ratio of ${\cal B}_{\mbox{\scriptsize min}}$ to ${\cal A}_{\mbox{\scriptsize min}}$ --- see (\ref{eq1.7}) and (\ref{eq1.8}) --- to the average of the $Q$-minima, $Q_{\mbox{\scriptsize min}}^{\pm}(T;L)$. As mentioned above, ${\cal A}_{\mbox{\scriptsize min}}$ and ${\cal B}_{\mbox{\scriptsize min}}$ vanish identically in the absence of the mixing coefficients, making the ratio ill-defined. On the other hand, when the system is asymmetric the ratio exhibits rather complex finite-size corrections in which both ${\cal A}_{\mbox{\scriptsize min}}$ and ${\cal B}_{\mbox{\scriptsize min}}$ contain terms varying as $L^{-\beta/\nu}$, $L^{-(\Delta -1)/\nu}$, etc. One realizes, therefore, that there is no {\em universal} finite-size scaling function which yields the diameter in a unique fashion --- in contrast to the case of the density discontinuity $\Delta\rho_\infty (T)$ \cite{kim:fis:lui,kim:fis2}. However, we demonstrate here that there are two extremal or limiting cases which yield {\em effective} universal scaling functions: The first arises when the pressure mixing is absent or so small as to be negligible: see (\ref{eq3.19}) below. The HCSW fluid belongs to this category as demonstrated in Sec.\ \ref{sec7.1}. The second case is found when the pressure-mixing term dominates over other contributions. An example will be the RPM electrolyte which then provides us with the other limiting scaling function: see (\ref{eq3.18}) and Sec.\ \ref{sec7.2}. Furthermore, as for the density discontinuity $\Delta\rho_{\infty}(T)$ \cite{kim:fis2}, these limiting scaling functions are analytically represented in (\ref{eq.Dmin_hcsw}) and (\ref{eq.Dmin_rpm}).

The balance of this article is organized as follows: In Sec.\ \ref{sec2} the scaling behavior of the $Q_{L}(\langle\rho\rangle_L)$ is presented on the basis of finite-size scaling theories. In Sec.\ \ref{sec3} the analysis of the $Q$-minima is presented. Section \ref{sec4} is devoted to estimation of the Yang-Yang ratios for the HCSW fluid and the RPM by using the critical order-parameter distribution of the Ising model and estimated nonuniversal amplitudes. Our best estimates for ${\cal R}_{\mu}$ are given in Eqs.\ (\ref{eq4.23}) and (\ref{eq4.26}). In Sec.\ \ref{sec5} the scaling algorithm for estimating the coexistence-curve diameter is presented. In Sec.\ \ref{sec6} an exactly soluble decorated lattice gas model is considered to construct one limiting case of the universal scaling function for the diameter. In Sec.\ \ref{sec7} we estimate the diameters of the HCSW fluid and the RPM electrolyte and compare them with previous results. Section \ref{sec8} summarizes the article and provides a brief discussion.

\section{\label{sec2}Finite-size scaling theories}

In this section we derive the scaling behavior of $Q_{L}(T,\langle\rho\rangle_{L})$ in order to provide the necessary theoretical background for estimating the Yang-Yang ratio, ${\cal R}_{\mu}$, and the coexistence-curve diameter, $\rho_{\mbox{\scriptsize diam}}(T)$. First, we consider the large $L$ behavior where one can obtain $Q_{L}(T,\langle\rho\rangle_{L})$ explicitly on the basis of the double-Gaussian approximation. We then study $Q_{L}(T,\langle\rho\rangle_{L})$ in the critical region below $T_{c}$ via the complete finite-size scaling theory \cite{kim:fis}.

\subsection{\label{sec2.1}Double-Gaussian limit}

For sufficiently large $L$ below $T_c$, we may study $Q_{L}(T,\langle\rho\rangle_{L})$ in the two-phase region on the basis of the two-Gaussian approximation for the probability distribution, ${\cal P}_{L}(\rho;T,\mu)$, of the density $\rho$ (at a fixed $T$ and $\mu$). This may be written
  \begin{eqnarray}
   {\cal P}_{L}(\rho;T,\mu) & \approx & C_{L}\{ \chi_{-}^{-1/2}\exp [-(\rho-\rho^{-})^{2}L^{d}/2\kb T\chi_{-}] \nonumber \\
  &  &  +\, \chi_{+}^{-1/2}\exp [-(\rho-\rho^{+})^{2}L^{d}/2\kb T\chi_{+}]\}  \nonumber \\
  &  &  \times \exp [\rho(\mu-\mu_{\sigma})L^{d}/\kb T],  \label{eq9}
  \end{eqnarray}
where $C_{L}(\mu,T)$ is a normalization constant while the $\chi_{\pm}(T)$ are the infinite-volume susceptibilities [defined via $\chi=(\partial\rho/\partial\mu)_{T}$] at $\rho=\rho^{\pm}(T)\pm$. It is then straightforward to calculate the mean density, $\langle\rho\rangle_{L}$, and various moments, $\langle m^{k}\rangle_{L}$, with $m=\rho-\langle\rho\rangle_{L}$, for $k=2,3,\cdots$, as functions of $T$ and $\mu$.

Following Ref.\ \cite{kim:fis}, let us introduce, for simplicity, the average and difference susceptibilities
  \begin{equation}
   \bar{\chi}(T) = \mbox{$\frac{1}{2}$}(\chi_{+}+\chi_{-}) \hspace{0.2in} \mbox{and} \hspace{0.2in} \chi_{0}(T) = \mbox{$\frac{1}{2}$}(\chi_{+}-\chi_{-}),  \label{eq10}
  \end{equation}
and a convenient parameter (related to the ordering field)
  \begin{equation}
   {\cal T}(T,h;L) = \tanh [h(\rho_{0}+\mbox{$\frac{1}{2}$}\chi_{0}h)L^{d}],  \label{eq11}
  \end{equation}
where $\rho_{0}(T)=\frac{1}{2}\Delta\rho_{\infty}(T)$ [see (\ref{eq.discon})] and the reduced field/chemical potential is defined via
  \begin{equation}
   h = [\mu-\mu_{\sigma}(T)]/\kb T.  \label{eq12}
  \end{equation}
The reduced dimensionless average and difference susceptibilities are then defined by
  \begin{eqnarray}
   X(L,T) & \equiv & \bar{\chi}(T)\kb T/\rho_{0}^{2}L^{d}, \nonumber \\
   X_{0}(L,T) & \equiv & \chi_{0}(T)\kb T/\rho_{0}^{2}L^{d}.  \label{eq13}
  \end{eqnarray}
Note that $X$ and $X_{0}$ approach zero when $L\rightarrow \infty$ since $\bar{\chi}$ and $\chi_0$ are finite below $T_c$.

After integrating (\ref{eq9}) multiplied by $\rho$, one may write the mean density as
  \begin{equation}
   \langle\rho\rangle_{L} = \rhodiam(T) + \bar{\chi}h + [\rho_{0}(T) + \chi_{0}h]{\cal T}.  \label{eq14}
  \end{equation}
The expression for $Q_{L}$ can then be written as a ratio of two polynomials of fourth order in $\langle\rho\rangle_{L}$ as presented in Ref.\ \cite{kim:fis}: see (4.19)-(4.25).

In the thermodynamic limit $(L\rightarrow\infty)$, $Q_{L}(T,\langle\rho\rangle_L )$ becomes zero at the coexistence curve boundary, ${\cal T}=\pm 1$. For large but finite $L$, however, we expect $Q_{L}(T,\langle\rho\rangle_L )$ to exhibit two isothermal minima close to zero at ${\cal T}_{\pm} = \pm 1 + \Delta{\cal T}_{\pm}$. To find $\Delta{\cal T}_{\pm}$ in terms of $X$ and $X_{0}$, we use (\ref{eq14}) to expand $Q_{L}$ in terms of $\bar{\chi}\propto X$ and $\chi_0 \propto X_0$ up to linear order to obtain
  \begin{equation}
   Q_{L}(T,\langle\rho\rangle_L ) = \frac{[2\Delta{\cal T}_{\pm}\pm X+ X_{0} + \cdots]^{2}}{4\Delta{\cal T}_{\pm}[2 + 3(X\mp X_{0}) + \cdots]}.  \label{eq15}
  \end{equation}
Solving the equation $(\partial Q_{L}/\partial{\cal T}) = 0$ for $\Delta{\cal T}_{\pm}$ then yields the minima at
  \begin{equation}
   \Delta{\cal T}_{\pm} = \pm \mbox{$\frac{1}{2}$}(X\pm X_{0}) + {\cal O}_{2},  \label{eq16}
  \end{equation}
where ${\cal O}_{2}$ represents terms of order $X^{i} X_{0}^{j}$ with $i+j\geq 2$. On substituting this into (\ref{eq15}), one finds
  \begin{equation}
   Q_{\mbox{\scriptsize min}}^{\pm} = X \pm X_{0} + {\cal O}_{2}.  \label{eq17}
  \end{equation}
Note that for the symmetric case (i.e., $X_0 = 0$) these results agree with those given in Ref.\ \cite{kim:fis} up to linear order in $X$. For the corresponding densities of these minima, we follow \cite{kim:fis:lui} and \cite{kim:fis2} and define the reduced density deviation by
 \begin{equation}
  y(T;\rho) \equiv 2[\rho-\rho_{\mbox{\scriptsize diam}}(T)]/\Delta\rho_{\infty}(T),  \label{eq18}
 \end{equation}
where $\rho_{\mbox{\scriptsize diam}}(T)$ is the diameter while $\Delta\rho_{\infty}(T)$ is defined in (\ref{eq.discon}); here and below we use $\rho$ to represent the mean density, $\langle\rho\rangle_L$, unless undesirable ambiguity arises. Note that $y$ takes the values $\pm 1$ at $\rho=\rho^{\pm}(T)$. Using (\ref{eq14}) and (\ref{eq16}) yields, after some algebra,
  \begin{equation}
   y_{\mbox{\scriptsize min}}^{\pm} = \pm [1 + \mbox{$\frac{1}{2}$} X\ln (4/eX) \pm \mbox{$\frac{1}{2}$}X_{0}\ln (4/e^{2} X)] + \cdots.  \label{eq19}
  \end{equation}

By taking the mean in (\ref{eq17}) and the difference in (\ref{eq19}), we obtain the {\em scaling relation}
 \begin{equation}
   \Delta y_{\mbox{\scriptsize min}} \equiv \mbox{$\frac{1}{2}$} (y_{\mbox{\scriptsize min}}^{+}-y_{\mbox{\scriptsize min}}^{-})  = 1 + \mbox{$\frac{1}{2}$}q + \cdots,  \label{eq20}
 \end{equation}
where
 \begin{equation}
  q\equiv \bar{Q}_{\mbox{\scriptsize min}}\ln[4/e\bar{Q}_{\mbox{\scriptsize min}}] \hspace{0.1in} \mbox{with} \hspace{0.1in} \bar{Q}_{\mbox{\scriptsize min}}\equiv \mbox{$\frac{1}{2}$}(Q_{\mbox{\scriptsize min}}^{+}+Q_{\mbox{\scriptsize min}}^{-}).  \label{eq21}
 \end{equation}
Similarly, from (\ref{eq1.7}), (\ref{eq1.8}), (\ref{eq17}) and (\ref{eq19}) one obtains 
 \begin{eqnarray}
  {\cal D}_{\mbox{\scriptsize min}} & \equiv & \frac{{\cal B}_{\mbox{\scriptsize min}}}{{\cal A}_{\mbox{\scriptsize min}}} = \mbox{$\frac{1}{2}$}\bar{q} + \cdots, \nonumber \\
  \bar{q} & \equiv & \bar{Q}_{\mbox{\scriptsize min}}\ln[4/e^{2}\bar{Q}_{\mbox{\scriptsize min}}]=q-\bar{Q}_{\mbox{\scriptsize min}}.  \label{eq22}
 \end{eqnarray}
Notice that these relations are {\em universal} up to the leading order in $q$ (or $\bar{q}$) and are asymptotically exact when $\bar{Q}_{\mbox{\scriptsize min}}\rightarrow 0$ (or $L\rightarrow\infty$).

\subsection{\label{fss}Finite-size scaling closer to criticality}

To understand the behavior of $Q_{L}(T,\rho)$ near criticality, we employ the complete scaling theory ({\ref{eq1.1})-(\ref{eq1.5}) which has been extended recently to finite systems of dimensions $L^{d}$ with periodic boundary conditions \cite{kim:fis}. The finite-size scaling ansatz asserts \cite{kim:fis,pri:fis}
  \begin{equation}
   \rho_c\tilde{p} \approx L^{-d}Y(x,z), \hspace{0.1in} x\equiv D_{L}\tilde{t}L^{1/\nu},\hspace{0.1in} z\equiv U_{L}\tilde{h}L^{\Delta/\nu},  \label{eq2.1}
  \end{equation}
in which $Y(x,z)$ is the basic scaling function while we have imposed the hyperscaling relation $d\nu = 2-\alpha$ (valid for $d<4$) and, for simplicity, neglected corrections to scaling. Here $\nu$ is the critical exponent for the correlation length, while $D_{L}$ and $U_{L}$ are {\em nonuniversal} amplitudes of dimensions $L^{-1/\nu}$ and $L^{-\Delta/\nu}$, respectively, which depend on the details of the system under consideration: see below. The scaling function $Y(x,z)$ is {\em universal} and thus independent of microscopic details of the system; but it depends on the geometry and the boundary conditions imposed. According to the underlying symmetry, $Y(x,z)$ is an even function of $z$.

Since the finite-size scaling function, $Y(x,z)$, is analytic and even in $z$, one can expand it as
 \begin{eqnarray}
   Y(x,z) & = & Y_{00} + Y_{10}x + Y_{20}x^2 + Y_{30}x^3 + \cdots \nonumber \\
  & & +\, z^2 (Y_{02} + Y_{12}x + Y_{22}x^2 + Y_{32}x^3 + \cdots) \nonumber \\
  & & +\, z^4 (Y_{04} + Y_{14}x + Y_{24}x^2 + Y_{34}x^3 + \cdots) \nonumber \\
  & & + \cdots.  \label{eq.Yexpand}
 \end{eqnarray}
One may further normalize $Y(x,z)$ by choosing the nonuniversal amplitudes, $D_{L}$ and $U_{L}$, so that one has $Y_{10}=Y_{02}=1$. The condition $Y_{02}=1$, in particular, will be used later for estimating the Yang-Yang ratio, ${\cal R}_{\mu}$.

To recover the bulk limit of the scaling form (\ref{eq1.5}), one may set $L=1/|D_{L}\tilde{t}|^{\nu}$ and formally let $L\rightarrow\infty$. This yields (\ref{eq1.5}) with the identifications
 \begin{eqnarray}
   Q=|D_{L}|^{2-\alpha}/\rho_c \hspace{0.2in} & \mbox{and} & \hspace{0.2in} U=U_{L}/|D_{L}|^{\Delta},  \label{eq.Q_U} \\
    W_{\pm}(z) & = & Y(\pm 1,z). \label{eq2.2}
  \end{eqnarray}
In the two-phase region ($\tilde{t}<0$), the scaling function $W_{-}(z)$ then has the expansion
  \begin{equation}
   W_{-}(z) = W_{-0} + W_{-1}|z| + W_{-2}z^{2} + \cdots.  \label{eq2.3}
  \end{equation}
This ensures the density discontinuity at the phase boundary.

The reduced dimensionless number density and susceptibility are conveniently defined by
  \begin{equation}
    \check{\rho} = \rho/\rho_c \equiv (\partial\check{p}/\partial\check{\mu})_{t}, \hspace{0.2in} \check{\chi}_{NN} \equiv (\partial^{2}\check{p}/\partial\check{\mu}^{2})_{t}.  \label{eq2.4}
  \end{equation}
Similarly one may define the generalized susceptibilities, $\check{\chi}_{N^{k}}\equiv (\partial^{k}\check{p}/\partial\check{\mu}^{k})$, $k=3$, $4$, $\cdots$, which will be used in the following section.

\subsection{\label{sec2.3}Finite-size scaling behavior of \boldmath $Q_{L}$}

To obtain $Q_{L}(T,\rho)$ in terms of the scaling variables $x\propto \tilde{t}L^{1/\nu}$ and $z\propto \tilde{h}L^{\Delta/\nu}$, we may notice that $Q_{L}$ can be expressed in terms of the generalized susceptibilities, $\check{\chi}_{N^{k}}$; namely, $Q_{L}$ is equivalent to $\rho_c V (\check{\chi}_{NN})^{2}/\check{\chi}_{N^{4}}$ where $V$ is the volume of the system. These susceptibilities then require the derivatives of the pressure, $p$, with respect to the chemical potential, $\mu$, at fixed $T$.

In order to compute the susceptibilities in terms of the scaling function $Y(x,z)$ in (\ref{eq2.1}), we may, first, obtain the reduced density $\check{\rho}$ by differentiating (\ref{eq2.1}) with respect to $\check{\mu}$ at fixed $T$. Using the scaling fields (\ref{eq1.1})-(\ref{eq1.3}) yields \cite{kim:fis}
  \begin{eqnarray}
    \check{\rho} & = & 1 + e_{1}A_{j}L^{-\kappa}[ (\partial_{z}Y) - j_{2}A_{j}L^{-\kappa}(\partial_{z}Y)^{2} \nonumber \\
  &  & - A_{l}L^{-\lambda}(\partial_{x}Y) + \cdots ],   \label{eq2.6}
  \end{eqnarray}
where, setting $l_{0}\equiv 1$ \cite{kim:fis:ork}, the exponents and nonuniversal amplitudes are \cite{kim2}
  \begin{eqnarray}
    e_{1} & = & 1-j_{2}, \hspace{0.2in} \kappa = \beta/\nu, \hspace{0.2in} \lambda = (\Delta -1)/\nu, \label{eq2.7} \\
    A_{j} & = & U_{L}/\rho_c, \hspace{0.3in} A_{l} = (l_{1}+j_{1})D_{L}/e_{1}U_{L},  \label{eq2.8}
  \end{eqnarray}
while, for simplicity, we have adopted the notations
  \begin{equation}
   (\partial_{x}Y) \equiv (\partial Y/\partial x), \hspace{0.2in} (\partial_{x}^{2}Y) \equiv (\partial^{2}Y/\partial x^{2}), \hspace{0.2in} \mbox{etc.}  \label{eq2.9}
  \end{equation}

Differentiating (\ref{eq2.6}) with respect to $\check{\mu}$ then yields the susceptibility $\check{\chi}_{NN}$ as
  \begin{eqnarray}
   \check{\chi}_{NN} & = & e_{1}^{2}\rho_c A_{j}^{2}L^{\gamma/\nu}[ (\partial_{z}^{2}Y) - 3j_{2} A_{j}L^{-\kappa}(\partial_{z}Y)(\partial_{z}^{2}Y) \nonumber \\
  &  & - 2A_{l} L^{-\lambda}(\partial_{x}\partial_{z}Y) + \cdots ]. \label{eq2.10}
  \end{eqnarray}
Finally, the fourth-order susceptibility is
  \begin{eqnarray}
   \check{\chi}_{N^{4}} & = & e_{1}^{4}\rho_c^{3}A_{j}^{4}L^{(\gamma+2\Delta)/\nu}[(\partial_{z}^{4}Y) \nonumber \\
  &  & - 5j_{2}A_{j}L^{-\kappa}\{ (\partial_{z}Y)(\partial_{z}^{4}Y) + 2(\partial_{z}^{2}Y)(\partial_{z}^{3}Y)\} \nonumber \\
  &  & - 4A_{l}L^{-\lambda}(\partial_{x}\partial_{z}^{3}Y) + \cdots ].  \label{eq2.12}
  \end{eqnarray}

The desired scaling form for $Q_{L}$ in the grand canonical representation (presented in Ref.\ \cite{kim:fis:lui} without any derivation) is then readily obtained from (\ref{eq2.10}) and (\ref{eq2.12}). After some algebra, we find the central result
  \begin{eqnarray}
   Q_{L}(T,\rho;L) & = & {\cal Q}_{Q}(x,z)[1+j_{2}A_{j}L^{-\kappa}{\cal Q}_{j}(x,z)\nonumber \\
  &  & + A_{l}L^{-\lambda}{\cal Q}_{l}(x,z) + \cdots ],  \label{eq2.13}
  \end{eqnarray}
where the {\em universal} scaling functions are given explicitly by
  \begin{eqnarray}
    {\cal Q}_{Q}(x,z) & = & (\partial_{z}^{2}Y)^{2}/(\partial_{z}^{4}Y),  \label{eq2.14} \\
    {\cal Q}_{j}(x,z) & = & -(\partial_{z}Y) + 10(\partial_{z}^{2}Y)(\partial_{z}^{3}Y)/(\partial_{z}^{4}Y),  \label{eq2.15} \\
    {\cal Q}_{l}(x,z) & = & 4[ (\partial_{x}\partial_{z}^{3}Y)/(\partial_{z}^{4}Y) - (\partial_{z}\partial_{x}Y)/(\partial_{z}^{2}Y)].   \label{eq2.16}
  \end{eqnarray}
Hence, the symmetry of $Y(x,z)$ implies that ${\cal Q}_{Q}(x,z)$ is even in $z$ while ${\cal Q}_{j}(x,z)$ and ${\cal Q}_{l}(x,z)$ are odd. The pressure-mixing coefficient $j_{2}$ thus yields the leading antisymmetric $L$-dependent correction term with a decay exponent $\kappa=\beta/\nu$.

\subsection{\label{sec2.4}Finite-size scaling behavior of the reduced density deviation, {\boldmath $y$}}

To derive the scaling form for the scale-free density deviation $y(T;L)$, as defined in (\ref{eq18}), we first note that the infinite-volume half-discontinuity, $\rho_{0}(T)=\frac{1}{2}\Delta\rho_{\infty}(T)$, and the coexistence-curve diameter vary asymptotically as
  \begin{eqnarray}
    \rho_{0}(T)/\rho_c & = & B_{0}|t|^{\beta} + \cdots, \label{eq2.17a} \\
    \rhodiam(T)/\rho_c & = & 1 + A_{2\beta}|t|^{2\beta}+A_{1-\alpha}|t|^{1-\alpha} + \cdots, \label{eq2.17}
  \end{eqnarray}
where the amplitudes are explicitly given by
  \begin{eqnarray}
    B_{0} & = & e_{1}QUW_{-1}|\tau|^{\beta},\hspace{0.1in} A_{2\beta} = -j_{2}B^{2}/e_{1}, \nonumber \\
   A_{1-\alpha} & = & (2-\alpha)(l_{1}+j_{1})QW_{-0}|\tau|^{1-\alpha},  \label{eq2.19}
  \end{eqnarray}
in which we have introduced the mixing factor \cite{kim:fis:ork}
 \begin{equation}
   \tau = 1-k_{1}l_{1}-(k_{0}+k_{1})(j_{1}+j_{2}l_{1})/e_{1},  \label{eq.tau}
 \end{equation}
while $Q$ and $U$ are related to $U_{L}$ and $D_{L}$ via the relations presented in (\ref{eq.Q_U}), and $W_{-0}$ and $W_{-1}$ are expansion coefficients of $W_{-}(z)$: see (\ref{eq2.3}). Note that, for simplicity, we have neglected many higher order terms including corrections to scaling \cite{kim:fis:ork}.

On using (\ref{eq2.6}), (\ref{eq2.17a})-(\ref{eq2.19}), one finds, after some algebra, that the reduced density deviation (\ref{eq18}) has the expansion
  \begin{eqnarray}
    y(T,\rho;L) & = & {\cal Y}(x,z)[ 1 + j_{2}A_{j}L^{-\kappa}{\cal Y}_{j}(x,z) \nonumber \\
  &  & + A_{l}L^{-\lambda}{\cal Y}_{l}(x,z) + \cdots],  \label{eq2.20}
  \end{eqnarray}
where the {\em universal} auxiliary scaling functions are
  \begin{eqnarray}
   {\cal Y}(x,z) & = & (\partial_{z}Y)/W_{-1}|x|^{\beta},  \label{eq2.21} \\
   {\cal Y}_{j}(x,z) & = & - (\partial_{z}Y) + W_{-1}^{2}|x|^{2\beta}/(\partial_{z}Y),  \label{eq2.22} \\
   {\cal Y}_{l}(x,z) & = & -\frac{ (\partial_{x}Y) +(2-\alpha)W_{-0}|x|^{1-\alpha}}{(\partial_{z}Y)}.  \label{eq2.23}
  \end{eqnarray}
Notice that in contrast to the scaling functions, ${\cal Q}_{Q}$, ${\cal Q}_{j}$ and ${\cal Q}_{l}$ in (\ref{eq2.14})-(\ref{eq2.16}), ${\cal Y}(x,z)$, ${\cal Y}_{j}(x,z)$ and ${\cal Y}_{l}(x,z)$ are singular at $x=0$. Furthermore, ${\cal Y}(x,z)$ diverges as $|x|^{-\beta}$ when $x\rightarrow 0$.

So far we have derived the scaling behavior of $Q_{L}(T,\rho;L)$ and $y(T,\rho;L)$ in terms of $x\propto \tilde{t}L^{1/\nu}$ and $z\propto \tilde{h}L^{\Delta/\nu}$ near criticality. However, our goal is to obtain the $Q$-minima and their locations, $Q_{\mbox{\scriptsize min}}^{\pm}(T;L)$ and $\rho_{\mbox{\scriptsize min}}^{\pm}(T;L)$, and to derive scaling relations between them. These expressions will then serve to estimate $j_{2}$ quantitatively and, likewise, the diameter near criticality.

\section{\label{sec3}Analysis of the {\boldmath $Q$}-minima}

\subsection{\label{sec3.1}{\boldmath $Q$}-minima and their locations}

From the observations of simulation data \cite{kim:fis,kim:fis:lui,bin:lan} and as borne out in the thermodynamic limiting form, $Q_{L}(T,\rho;L)$ is expected to exhibit two isothermal minima near the coexistence-curve boundary. For symmetric cases (i.e., $l_{1}=j_{1}=j_{2}=0$), it is then obvious that ${\cal Q}_{Q}(x,z)$ in (\ref{eq2.14}) must have equi-height minima at some value of $z$, say $\pm z_{\mbox{\scriptsize m}}(x)$, for fixed $x$. Owing to the analyticity of the finite-size scaling function $Y(x,z)$, the function $z_{\mbox{\scriptsize m}}(x)$ is also analytic in $x$.

Now we expand the scaling function $Y(x,z)$ about these minima at $\pm z_{\mbox{\scriptsize m}}(x)$ and, by virtue of the symmetry of $Y(x,z)$, obtain
  \begin{equation}
   Y(x,z) = \sum_{i=0}^{\infty}(\pm)^{i}a_{i}(x)[z\mp z_{\mbox{\scriptsize m}}(x)]^{i},  \label{eq3.1}
  \end{equation}
where the $a_{i}(x)$ are {\em universal} expansion coefficients of $Y$. Since $Y(x,z)$ is analytic everywhere for finite $L$, the $a_{i}(x)$ must be also analytic and have the {\em universal} expansions
  \begin{equation}
    a_{i}(x) = a_{i0} + a_{i1}x + a_{i2}x^{2} + \cdots.  \label{eq3.2}
  \end{equation}
From (\ref{eq2.13}) and (\ref{eq2.14}), in the symmetric case, we then find
  \begin{equation}
   Q_{\mbox{\scriptsize min}}^{\pm}(T;L) \approx Q_{\mbox{\scriptsize min}}^c [ 1+ (2a_{21}-a_{41})x + \cdots] \equiv {\cal Q}_{\mbox{\scriptsize m}}(x),  \label{eq3.3}
  \end{equation}
where $Q_{\mbox{\scriptsize min}}^c = a_{20}^{2}/a_{40}$ is a universal constant.

In the presence of the mixing coefficients $l_{1}$, $j_{1}$ and $j_{2}$, the locations of the minima will be shifted by amounts, say $\Delta z_{\pm}(x)$. To find these shifts, we solve the equation $(\partial Q_{L}/\partial z)=0$ for $\Delta z_{\pm}$ perturbatively. After some algebra, one obtains
  \begin{equation}
    \Delta z_{\pm}(x) = [ j_{2}A_{j}L^{-\kappa}b_{1}(x) + A_{l}L^{-\lambda}b_{2}(x)]/b_{0}(x) + \cdots,  \label{eq3.4}
  \end{equation}
where $\kappa$, $\lambda$, $A_{j}$ and $A_{l}$ were introduced in (\ref{eq2.7}) and (\ref{eq2.8}) while the universal scaling functions are
  \begin{eqnarray}
    b_{0}(x) & = & 2 \frac{a_{4}(x)}{a_{2}(x)} - \frac{a_{6}(x)}{a_{4}(x)} + 2\left( \frac{a_{3}(x)}{a_{2}(x)}\right)^{2},  \label{eq3.5}  \\
    b_{1}(x) & = & -9a_{2}(x) + 10 \left(a_{3}(x)\right)^{2}/a_{4}(x),  \label{eq3.6} \\
    b_{2}(x) & = & 4 \left[ \frac{a_{2}^{\prime}}{a_{2}}-\frac{a_{4}^{\prime}}{a_{4}}-\frac{a_{1}^{\prime}a_{3}}{a_{2}^{2}} + 2\frac{a_{3}^{\prime}a_{3}}{a_{4}a_{2}}\right],  \label{eq3.7}
  \end{eqnarray}
in which $a_{i}^{\prime}(x)\equiv da_{i}/dx$.

Substituting these into (\ref{eq2.13})-(\ref{eq2.16}) via (\ref{eq3.1}) finally yields
  \begin{equation}
   Q_{\mbox{\scriptsize min}}^{\pm}(T;L) = {\cal Q}_{\mbox{\scriptsize m}}(x)[ 1 \pm j_{2}A_{j}c_{j}(x)L^{-\kappa} \pm A_{l}c_{l}(x)L^{-\lambda} + \cdots ],  \label{eq3.8}
  \end{equation}
where the auxiliary scaling functions, again universal, are
  \begin{eqnarray}
   c_{j}(x) & = & -a_{1}(x) + 10 a_{2}(x)a_{3}(x)/a_{4}(x),  \label{eq3.9} \\
   c_{l}(x) & = & 4[a_{3}^{\prime}(x)/a_{4}(x)-a_{1}^{\prime}(x)/a_{2}(x)].  \label{eq3.10}
  \end{eqnarray}
Note that the $b_{j}(x)$ in (\ref{eq3.5})-(\ref{eq3.7}) do not enter here. The basic asymmetry factor, ${\cal A}_{\mbox{\scriptsize min}}$, defined in (\ref{eq1.7}) is thus given explicitly by
  \begin{equation}
   {\cal A}_{\mbox{\scriptsize min}}(T;L) = j_{2}A_{j}c_{j}(x)L^{-\kappa} + A_{l}c_{l}(x)L^{-\lambda} + \cdots.  \label{eq3.11}
  \end{equation}
As mentioned in the Introduction, we see that the dominant contribution to ${\cal A}_{\mbox{\scriptsize min}}$ arises from the pressure-mixing coefficient $j_{2}$ with a decay exponent $\kappa=\beta/\nu$.

On the other hand, the locations of the minima, $y_{\mbox{\scriptsize min}}^{\pm}(T;L)$, can be obtained similarly by using (\ref{eq2.20})-(\ref{eq2.23}) and (\ref{eq3.4}). This yields
  \begin{eqnarray}
   y_{\mbox{\scriptsize min}}^{\pm}(T;L) & = & \pm {\cal Y}_{\mbox{\scriptsize m}}(x)[1\pm j_{2}A_{j}d_{j}(x)L^{-\kappa} \nonumber \\
  &  & \pm A_{l}d_{l}(x)L^{-\lambda} + \cdots],   \label{eq3.12}
  \end{eqnarray}
where the scaling functions are
  \begin{eqnarray}
   {\cal Y}_{\mbox{\scriptsize m}}(x) & = & a_{1}(x)/|x|^{\beta}W_{-1}, \label{eq3.13} \\
   d_{j}(x) & = & -a_{1} + \frac{a_{2}b_{1}}{a_{1}b_{0}} + \frac{W_{-1}^{2}|x|^{2\beta}}{a_{1}}, \label{eq3.14} \\
   d_{l}(x) & = & -\frac{a_{0}^{\prime}}{a_{1}} + \frac{a_{2}b_{2}}{a_{1}b_{0}} - (2-\alpha)\frac{W_{-0}|x|^{1-\alpha}}{a_{1}},  \label{eq3.15}
  \end{eqnarray}
while the $b_{j}(x)$ are defined in (\ref{eq3.5})-(\ref{eq3.7}). Note that ${\cal Y}_{\mbox{\scriptsize m}}(x)$ diverges at criticality as $|x|^{-\beta}$ while $d_{j}(x)$ and $d_{l}(x)$ approach constants singularly as $|x|^{2\beta}$ and $|x|^{1-\alpha}$, respectively, when $x\rightarrow 0$.

\subsection{\label{sec3.2}Universal relations for the {\boldmath $Q$}-minima}

The scaling algorithm \cite{kim:fis:lui,kim:fis2}, which was used to estimate the density discontinuity, $\Delta\rho_{\infty}(T)$, requires a scaling relation between $\Delta y_{\mbox{\scriptsize min}}$ and $\bar{Q}_{\mbox{\scriptsize min}}$. From (\ref{eq3.8}) and (\ref{eq3.12}), one has
 \begin{equation}
   \Delta y_{\mbox{\scriptsize min}} = {\cal Y}_{\mbox{\scriptsize m}}(x) + \cdots, \hspace{0.1in} \bar{Q}_{\mbox{\scriptsize min}} = {\cal Q}_{\mbox{\scriptsize m}}(x) + \cdots. \label{eq3.16}
 \end{equation}
By formally solving for $x$ in terms of $\bar{Q}_{\mbox{\scriptsize min}}$ using (\ref{eq3.3}) and substituting into the first member of (\ref{eq3.16}), one obtains a function $\Delta y_{\mbox{\scriptsize min}}(\bar{Q}_{\mbox{\scriptsize min}})$. Note that its asymptotic behavior for $\bar{Q}_{\mbox{\scriptsize min}}\rightarrow 0$ follows from (\ref{eq20}).

The coexistence-curve diameter, $\rho_{\mbox{\scriptsize diam}}(T)$, can be estimated similarly via the scaling algorithm. For this purpose, we consider ${\cal D}_{\mbox{\scriptsize min}}$, defined in (\ref{eq22}), since ${\cal B}_{\mbox{\scriptsize min}}$ contains the diameter only. One should, however, notice that ${\cal D}_{\mbox{\scriptsize min}}$ is defined only for {\em asymmetric} cases. Using (\ref{eq3.8}) and (\ref{eq3.12}) then yields
 \begin{equation}
  {\cal D}_{\mbox{\scriptsize min}} = \frac{j_{2}A_{j}d_{j}(x)L^{-\kappa} + A_{l}d_{l}(x)L^{-\lambda}+\cdots}{j_{2}A_{j}c_{j}(x)L^{-\kappa}+A_{l}c_{l}(x)L^{-\lambda}+\cdots}.  \label{eq3.17}
 \end{equation}
Unlike $\Delta y_{\mbox{\scriptsize min}}$, ${\cal D}_{\mbox{\scriptsize min}}$ has a complicated structure of finite-size terms due to the mixing coefficients. One can clearly identify two limiting cases. First, when pressure-mixing is dominant, one simply has
 \begin{equation}
   {\cal D}_{\mbox{\scriptsize min}} = \frac{d_{j}(x)}{c_{j}(x)} + O(L^{-\lambda+\kappa}) \approx e_{j}(x).  \label{eq3.18}
 \end{equation}
Eliminating $x$ between this relation and (\ref{eq3.16}) and (\ref{eq3.3}) as before then yields an asymptotically universal relation between ${\cal D}_{\mbox{\scriptsize min}}$ and $\bar{Q}_{\mbox{\scriptsize min}}$. On the other hand, if pressure-mixing is absent or can be neglected, the form (\ref{eq3.17}) reduces to
 \begin{equation}
   {\cal D}_{\mbox{\scriptsize min}} = \frac{d_{l}(x)}{c_{l}(x)} + \cdots \approx e_{l}(x).  \label{eq3.19}
 \end{equation}
These results will play the central role in the subsequent analysis. Note, however, that when $\bar{Q}_{\mbox{\scriptsize min}}\rightarrow 0$ one recaptures the universal relation (\ref{eq22}) regardless of the particular mixing situation.

\section{\label{sec4}Detecting Yang-Yang anomalies}

In the previous section, the asymmetry factor, ${\cal A}_{\mbox{\scriptsize min}}(T;L)$, derived from the $Q$-minima was analyzed and it was shown that the pressure-mixing coefficient $j_{2}$ provides a leading term varying as $L^{-\beta/\nu}$ which dominates over other contributions: see Table \ref{tab1} for the Ising values for the universal exponents used in the following sections. In particular, at criticality $(x\propto\tilde{t}L^{1/\nu}=0)$ the result (\ref{eq3.11}) leads to
  \begin{equation}
   {\cal A}_{\mbox{\scriptsize min}}^c = j_{2} A_{j}\bar{c}_{j}L^{-\kappa} + A_{l}\bar{c}_{l}L^{-\lambda} + \cdots,  \label{eq4.1}
  \end{equation}
where the universal constants are
  \begin{eqnarray}
   \bar{c}_{j} & = & -a_{10} + 10a_{20}a_{30}/a_{40},  \label{eq4.2} \\
   \bar{c}_{l} & = & -4 a_{11}/a_{20} + 4 a_{31}/a_{40}.  \label{eq4.3}
  \end{eqnarray}
in which $a_{ij}$ are the coefficients of $x^{j}$ in the expansion of $Y(x,z)$ at the minima, $z=z_{\mbox{\scriptsize m}}$: see (\ref{eq3.2}) above. On the other hand, $A_{j}$ is a {\em nonuniversal} constant introduced in (\ref{eq2.8}), while $A_{l}\propto (l_{1}+j_{1})/(1-j_{2})$. Thus to measure, in particular, the pressure-mixing coefficient, $j_{2}$, one requires not only the simulation data for ${\cal A}_{\mbox{\scriptsize min}}^c$ but {\em also} information concerning the universal constant $\bar{c}_{j}$ and the nonuniversal amplitude $A_{j}$. 

\subsection{\label{sec4.1}Estimation of the universal constant \boldmath $\bar{c}_{j}$}

To determine the constant $\bar{c}_{j}$ as given by (\ref{eq4.2}), one must determine the expansion coefficients, $a_{j0}$, of the scaling function $Y(x,z)$ at $x=0$ about its minima: see (\ref{eq3.1}). Invoking universality, we thus consider a simple cubic ferromagnetic Ising model, to which universality class normal fluids are believed to belong. In this case the ``pressure-field'', $\rho_c\tilde{p}$, in (\ref{eq2.1}) corresponds to the reduced free energy density for a system of volume $L^{d}$ while the ordering field $\tilde{h}$ corresponds to the reduced dimensionless magnetic field. If we now define a scaled fluctuating magnetization density by $w\equiv A_{L}mL^{\beta/\nu}$ (where $m$ corresponds to the fluctuating reduced magnetization density) with the identification $A_{L}=1/U_{L}$, we find from (\ref{eq2.1}) that
  \begin{equation}
    \langle w^{k} \rangle = \langle (A_{L}mL^{\beta/\nu})^{k}\rangle = (\partial^{k} Y/\partial z^{k}).  \label{eq4.4}
  \end{equation}
Thus the expansion coefficients $a_{j0}$ can be obtained from the $j$-th moment of the scaled magnetization density at criticality.

Now in order to estimate the $\langle w^{k}\rangle$ numerically, we will utilize the most reliable and precise estimate available for the {\em universal} probability distribution function, $P_{L}^c (m)$, of the magnetization density, $m$, at criticality: this has been obtained numerically from careful simulations \cite{tsy:blo}. This function may be written in scaling form as
  \begin{equation}
   P_{L}^c (m) = C \exp\{ -\tilde{f}(w) + wz\}, \hspace{0.3in} w= A_{L}mL^{\beta/\nu},  \label{eq4.5}
  \end{equation}
where the nonuniversal amplitude $A_{L}$ is chosen so that the distribution has unit variance in $w$, while $C$ is a normalization constant; but one should notice from (\ref{eq.Yexpand}) and (\ref{eq4.4}) that this is equivalent to the normalization condition $Y_{02}=1$ for $Y(x,z)$: see Sec.\ \ref{fss}. Recall that $z=U_{L}\tilde{h}L^{\Delta/\nu}$ with $U_{L}=1/A_{L}$. Here $\tilde{f}(w)$ corresponds to a universal {\em canonical} scaling function. Tsypin and Bl\"{o}te \cite{tsy:blo} present this function approximately in the form
  \begin{equation}
    \tilde{f}(w) = \left[(w/w_{0})^{2}-1\right]^{2}\left[a (w/w_{0})^{2}+c\right].  \label{eq4.6}
  \end{equation}
where $w_{0}$, $a$ and $c$ are constants estimated via the simulations as
  \begin{equation}
   w_{0} = 1.134(4), \hspace{0.2in} a = 0.158(2), \hspace{0.2in} c = 0.776(2), \label{eq4.7}
  \end{equation}
where the uncertainties in parentheses refer to the last decimal place quoted \cite{tsy:blo}. Furthermore, the condition of unit variance yields $A_{L}\simeq 0.802$.

Using these forms, one may calculate the derivatives of $Y(0,z)$, i.e., $(\partial^{k} Y/\partial z^{k})$, numerically by computing the $k$-th moments of $w$. It is then straightforward to calculate $Q_{L}(T_c,\langle\rho\rangle_L )$ as a function of $z$: see Fig.\ \ref{fig1}. 
\begin{figure}[ht]
\vspace{-0.9in}
\centerline{\epsfig{figure=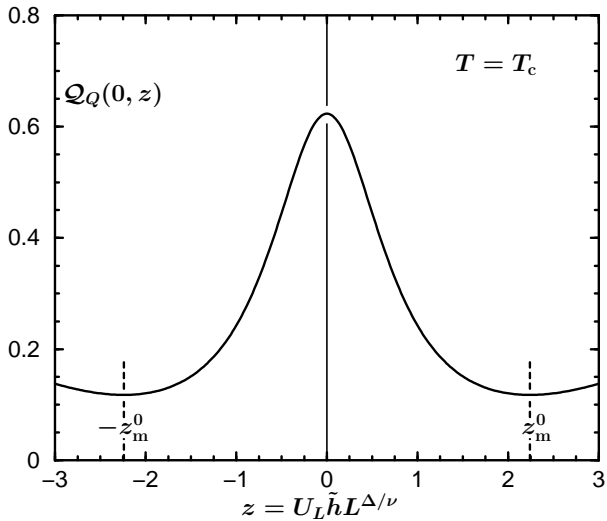,width=3.8in,angle=0}}
\vspace{-1.3in}
\caption{Plot of the universal critical-point scaling function $Q_{L}(T_{c};\langle\rho\rangle_L )\approx {\cal Q}_{Q}(0,z)$ vs $z=U_{L}\tilde{h}L^{\Delta/\nu}$ for the $(d=3)$ Ising model obtained numerically from (\ref{eq4.5}) \cite{tsy:blo}. The dashed lines locate the minima at $\pm z_{\mbox{\scriptsize m}}^{0}\simeq \pm 2.2395$. \label{fig1}}
\end{figure}
It exhibits two symmetric minima, at $z = \pm z_{\mbox{\scriptsize m}}^{0}$ where
  \begin{equation}
    Q_{\mbox{\scriptsize min}}^c = 0.1178(15) \hspace{0.3in} \mbox{and} \hspace{0.3in} z_{\mbox{\scriptsize m}}^{0}= 2.24(3).  \label{eq4.8}
  \end{equation}
Note, however, that $Q_{\mbox{\scriptsize min}}^c$ is about $7\%$ higher than the value $(\simeq 0.1107)$ found for the HCSW fluid via grand canonical simulations \cite{kim:fis:lui,kim5}. The universal constant, $Q^c$, at criticality (i.e., $z=0$) agrees with the estimated value for the Ising universality class, $Q^c \simeq 0.6236$ \cite{blo:lui:her,lui:fis:pan}. Calculating $\langle w^{k}\rangle$ at the minimum, $z=z_{\mbox{\scriptsize m}}^{0}$, then yields the desired expansion coefficients as
  \begin{eqnarray}
   a_{10} & \simeq & 1.2213, \hspace{0.2in} a_{20}\simeq 0.1751, \hspace{0.2in} a_{30} \simeq -0.1271, \nonumber \\
  & & \hspace{0.2in} a_{40}\simeq 0.2603, \hspace{0.2in} a_{60}\simeq 0.9878.  \label{eq4.9}
  \end{eqnarray}
Notice that in (\ref{eq3.3}) one obtains $Q_{\mbox{\scriptsize min}}^c =a_{20}^{2}/a_{40}\simeq 0.1178$ which is fully consistent with the estimate (\ref{eq4.8}). Finally, the universal constant $\bar{c}_{j}$ in (\ref{eq4.2}) is found to be
  \begin{equation}
    \bar{c}_{j}\simeq -2.077.  \label{eq4.10}
  \end{equation}

In reality the distribution of the magnetization, $P_{L}^c (m)$, should behave like $\exp(-c_{\delta} |m|^{\delta+1})$ for large $m$ with $\delta=\Delta/\beta\simeq 4.8$ for $(d=3)$ Ising universality class: see Table \ref{tab1}. Hence, the approximate form (\ref{eq4.6}) does not describe the behavior of the distribution properly for large $m$, even though $\delta +1\simeq 5.8$ is close to the exponent 6 embodied in this form. One can improve the approximation of Tsypin and Bl\"{o}te in order to capture the correct behavior of large $m$. However, we find that appropriate modifications leave the central estimate (\ref{eq4.10}) for $\bar{c}_{j}$ unchanged.

\subsection{\label{sec4.2}Nonuniversal amplitude \boldmath $U_{L}$}

The nonuniversal amplitude $U_{L}$ determining the value of $A_{j}$ via (\ref{eq2.8}) depends on the microscopic details of the system. Unlike the universal constant $\bar{c}_{j}$ obtained in the previous section, one must estimate $U_{L}$ separately for each model. Here we present a method for evaluating $U_{L}$ from knowledge of the leading nonuniversal critical amplitudes for bulk thermodynamic quantities.

In bulk near-critical systems, the susceptibility above $T_c$ and the order parameter below $T_c$ vary as \cite{liu:fis,fis:zin}
  \begin{eqnarray}
   M(T) & = & (\partial \bar{p}/\partial \bar{\mu}) \approx B |t|^{\beta},  \label{eq4.11} \\
   \chi_{NN}(T) & = & (\partial^{2}\bar{p}/\partial \bar{\mu}^{2}) \approx C^{+}t^{-\gamma},  \label{eq4.12}
  \end{eqnarray}
with $\bar{p}=p/k_{\mbox{\scriptsize B}}T$ and $\bar{\mu}=\mu/k_{\mbox{\scriptsize B}}T$, where the nonuniversal amplitudes can be obtained from the scaling formulation (\ref{eq1.5}) as \cite{kim:fis:ork}
  \begin{eqnarray}
   B & \equiv & \rho_{c}B_0 = (1-j_{2}) \rho_c QU W_{-1} |\tau|^{\beta},  \label{eq4.13} \\
   C^{+} & = & (1-j_{2})^{2}\rho_c QU^{2} W_{+2}|\tau|^{-\gamma}, \label{eq4.14}
  \end{eqnarray}
while $\tau$ is defined in (\ref{eq.tau}). Note that these amplitudes have dimensions $L^{-d}$. Recall that $W_{-1}$ and $W_{+2}$ are expansion coefficients of the scaling functions $W_{\pm}(z)$: see (\ref{eq2.2}) and (\ref{eq2.3}). Thus they are universal. Likewise note that $Q$ and $U$ are related to the finite-size amplitudes, $D_{L}$ and $U_{L}$, via (\ref{eq.Q_U}).

Now we may combine (\ref{eq4.13}) and (\ref{eq4.14}) in order to solve for $U_{L}$ in terms of $B$ and $C^{+}$. After some algebra, one finds
  \begin{equation}
    U_{L} = U D_{L}^{\Delta} = K \left[B^{\gamma}(C^{+})^{\beta}\right]^{1/(2-\alpha)}/|1-j_{2}|,  \label{eq4.15}
  \end{equation}
where $K$ is a universal constant given by
  \begin{equation}
    K = \left( W_{-1}^{\gamma}W_{+2}^{\beta}\right)^{-1/(2-\alpha)},  \label{eq4.16}
  \end{equation}
which can be obtained numerically from the well-studied Ising model. As mentioned above, we first notice that the condition of unit variance and the numerical simulations \cite{tsy:blo} yield $U_{L}=1/A_{L}\simeq 1/0.802$ for the $(d=3)$ nearest-neighbor simple cubic (sc) Ising model \cite{tsy:blo}: see (\ref{eq4.5}). The amplitudes $B$ and $C^+$ for this model have been studied and are given by \cite{liu:fis,kim6}
  \begin{equation}
    B = 1.66(3), \hspace{0.3in} C^{+} = 1.095(10).  \label{eq4.17}
  \end{equation}
From (\ref{eq4.15}) and (\ref{eq4.17}) together with the exponent values in Table \ref{tab1}, one then obtains
  \begin{equation}
   K = 0.881(12).  \label{eq4.18}
  \end{equation}

Now, one clearly sees from (\ref{eq4.15}) that to estimate $U_{L}$ for model fluids, information concerning the amplitudes $B$ and $C^{+}$ is crucial.

\subsection{\label{sec4.3}Estimation of the Yang-Yang ratios}

In this section we finally obtain estimates for the pressure-mixing coefficients $j_{2}$, and thereby the Yang-Yang ratios, ${\cal R}_{\mu}$, for the HCSW fluid and the RPM by utilizing the results obtained. In Fig.\ \ref{fig2} the asymmetry factors, ${\cal A}_{\mbox{\scriptsize min}}(L;T_c)$, at the critical temperature for the two models are presented as functions of $L^{-(\Delta-1)/\nu}$ and $L^{-\beta/\nu}$, respectively. 
\begin{figure}[ht]
\vspace{-0.3in}
\centerline{\epsfig{figure=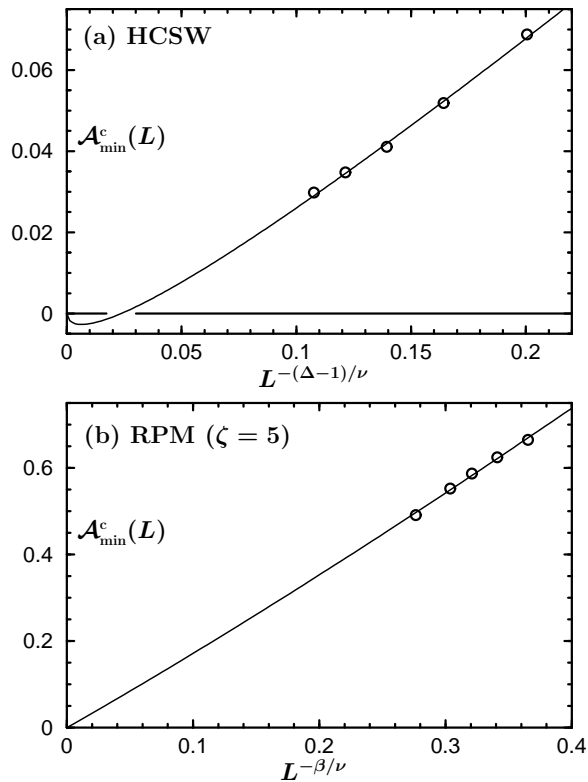,width=3.8in,angle=0}}
\vspace{-0.5in}
\caption{Plots of the asymmetry factors, ${\cal A}_{\mbox{\scriptsize min}}^{\mbox{\scriptsize c}}(L)$, as obtained from simulations for (a) the HCSW fluid (vs $L^{-(\Delta -1)/\nu}$) and (b) for the RPM (vs $L^{-\beta/\nu}$) with the exponent values listed in Table \ref{tab1}. The solid curves are fits of the data to (\ref{eq4.1}). \label{fig2}}
\end{figure}
If a Yang-Yang anomaly is present, the data should decay asymptotically as $L^{-\beta/\nu}$ when $L\rightarrow\infty$. The plot for the RPM clearly suggests that this highly asymmetric model has a nonvanishing Yang-Yang anomaly. On the other hand, the HCSW fluid would seem to have a quite small, if any, Yang-Yang anomaly. The solid lines are fits of the data to the formula (\ref{eq4.1}) neglecting the higher order terms. [See (\ref{eq4.22}) and (\ref{eq4.25}) below.] Note that the fit for the HCSW fluid exhibits a small negative leading amplitude. Based on these data --- the origin of which we first review briefly --- and the resulting fits we now describe the procedures used to estimate the Yang-Yang ratios, ${\cal R}_{\mu}$, quantitatively.

\subsubsection{\label{sec4.3.0}Details of simulations}

The asymmetry factors presented in Fig.\ \ref{fig2} and all the simulation data in the subsequent figures have been obtained via grand canonical Monte Carlo (GCMC) simulations in a cubic box of volume $V=L^3$. In GCMC one performs a simulation at a thermodynamic state point (SP) characterized by a given value of the temperature, $T$, and the chemical potential, $\mu$. In order to capitalize on the widely used multi-histogram reweighting technique \cite{fer:swe}, one measures the joint histogram of the fluctuating energy $U$ and the number of particles $N$ for different SPs \cite{ork:fis:pan,lui:fis:pan}. This approach enables one to extract the maximum amount of information from the simulations.

To obtain sufficiently accurate simulation data for the HCSW fluid, each simulation was performed for a total run length in the range of $20$-$80$$\times$$10^6$ MC steps, depending on the particular system size under investigation \cite{ork:fis:pan}. A total of 30-100 SPs (which broadly cover the critical region and the coexistence curve) have been used in the computations. Statistical uncertainties for the density $\rho=N/V$ and for the leading moments for each histogram are found to be less than $\frac{1}{2}\%$. On the other hand, for the RPM the total run length needed for each simulation is considerably larger owing to the much lower critical temperature, and to the great number of SPs necessary to ensure accurate data. For example, a total of 167 SPs were used for $L = 12a$ (where $a$ is the particle diameter), in which a typical SP has a $(2$-$10)$$\times$$10^4$ {\em independent} samples amounting to $\sim$$10^{10}$ MC steps \cite{lui:fis:pan}. For smaller systems, the number of independent samples employed is generally larger by factors of $10$-$50$. For each SP, the statistical uncertainties of the raw data are less than 1-2$\%$.

We know of no definitive or systematic study concerning the propagation of errors in the multi-histogram reweighting process. However, experience shows that when the SPs are closely spaced in regions of rapid changes in the computed quantities that encompass desired values of $T$ and $\mu$, the systematic uncertainties in the leading moments do not exceed and may well be appropriately less than the statistical uncertainties that characterize each SP. Hence, we believe that the errors associated with the data presented in Fig.\ \ref{fig2} and the subsequent figures are no larger than the symbol sizes (or, in plots like Fig.\ \ref{fig3}, no more than a couple of line thicknesses). Furthermore, the confirmation of a smooth and systematic variation of specific plots with increasing values of $L$, as seen in Fig.\ \ref{fig3}, serves as an important cross-check. When apparent erratic behavior is seen, larger runs and/or further SPs have been employed as a check.

\subsubsection{\label{sec4.3.1}Hard-core square-well fluid}

The HCSW fluid investigated here consists of hard-spheres of diameter $a$ with an attractive square-well of depth $\epsilon$ in the range $a\leq r\leq 1.5a$, where $r$ is the interparticle distance. Previously Orkoulas {\em et al.} \cite{ork:fis:pan} tried to estimate ${\cal R}_{\mu}$ via Monte Carlo (MC) simulations by computing directly the second derivative of the chemical potential along the phase boundary; they concluded that there was a negative but small Yang-Yang ratio, ${\cal R}_{\mu}\simeq -0.08$. Owing to the finite-size effects, however, they could not rule out a vanishing ${\cal R}_{\mu}$. Here we use the asymmetry factor, ${\cal A}_{\mbox{\scriptsize min}}$, at criticality as illustrated in Fig.\ \ref{fig2} to calculate ${\cal R}_{\mu}$ more precisely. In doing so, as mentioned above, one needs to compute $U_{L}$ which requires knowledge of the order parameter (or density-half-discontinuity), $\rho_{0}(T)$, and the susceptibility, $\chi_{NN}(T)$, above $T_c$.

Near criticality the appropriate expansions are
  \begin{eqnarray}
   \rho_{0}(T) & = & B|t|^{\beta}[1+b_{\theta}|t|^{\theta}+\cdots], \label{eq4.19a} \\
   \chi_{NN}(T) & = & C^{+}t^{-\gamma}[1+c_{\theta}t^{\theta}+\cdots],  \label{eq4.19} 
  \end{eqnarray}
where $\theta$ is the leading correction-to-scaling exponent. For $(d=3)$ Ising criticality one has $\theta\simeq 0.52$ \cite{gui:zin}, while $b_{\theta}$ and $c_{\theta}$ are nonuniversal amplitudes. Recent precise simulation of the coexisting densities for this HCSW fluid via the $Q$-minima scaling algorithm (see below) yields a reliable estimate for $B$: see Fig.\ 1 in Ref.\ \cite{kim:fis:lui}. On the other hand, to determine $C^{+}$, we have computed the finite-size susceptibility $\chi_{NN}(T;L)$ by simulations on the critical isochore $\rho=\rho_c$. In the bulk limit $\chi_{NN}(T)|t|^{\gamma}$ should then approach $C^{+}$ when $t\rightarrow 0$: see Fig.\ \ref{fig3}(a). 
\begin{figure}[ht]
\vspace{-0.4in}
\centerline{\epsfig{figure=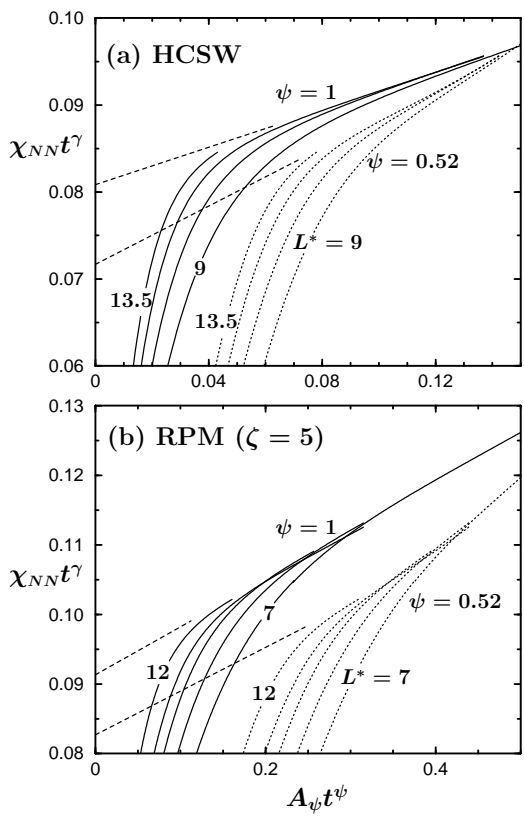,width=4.0in,angle=0}}
\vspace{-0.5in}
\caption{Plot for estimating the susceptibility amplitude $C^+$ from data for $\chi_{NN}$ on the critical isochore \cite{ork:fis:pan,lui:fis:pan}. Note that $t=(T-T_{c})/T_{c}$ while the values $\psi=1$ (solid lines) and $\psi=\theta=0.52$ (dotted lines) have been chosen for the auxiliary extrapolation exponent (together with, purely for clarity of presentation, $A_{1}=1$ and $A_{\theta}=0.4,\, 0.8$). For (a) the HCSW fluid and (b) the RPM electrolyte, the estimated critical densities are $\rho_c^\ast =\rho_{c}a^{3}=0.3206$ and $0.079$, respectively, while $L^{\ast}=L/a=9,10.5,12,13.5$ for the HCSW fluid, and $7,8,9,10,12$ for the RPM. The dashed lines represent approximate upper and lower extremal estimates for $C^{+}$.\label{fig3}}
\end{figure}
For finite systems, however, this product will approach zero when $t\rightarrow 0$, as seen in Fig.\ \ref{fig3}, since $\chi_{NN}(T;L)$ is always bounded for $L<\infty$. Nevertheless, we can estimate $C^{+}$ reasonably well by extrapolating the data of \cite{kim:fis:lui} to $t=0$. We conclude
  \begin{equation}
    Ba^3 = 0.612 \pm 0.005,  \hspace{0.1in} C^{+}a^3 = 0.076\pm 0.005, \hspace{0.1in} \mbox{(HCSW).}  \label{eq4.20}
  \end{equation}
Furthermore, the Ising fit to the $\rho_{0}(T)$ for the HCSW fluid via (\ref{eq4.19a}) with a further correction term varying as $|t|^{2\beta}$ \cite{kim:fis:ork} yields the leading correction-to-scaling amplitude as $b_{\theta}\simeq -1.04$. Using (\ref{eq4.15}) and (\ref{eq4.18}) now yields the HCSW nonuniversal amplitude as
  \begin{equation}
   U_{L}|1-j_{2}|a^{\Delta/\nu} = 0.410 \pm 0.012.  \label{eq4.21}
  \end{equation}

The fit of the ${\cal A}_{\mbox{\scriptsize min}}$ data to (\ref{eq4.1}), illustrated in Fig.\ 2, yields
  \begin{equation}
    j_{2}A_{j}\bar{c}_{j} \simeq -0.117, \hspace{0.1in} A_{l}\bar{c}_{l} \simeq 0.570, \hspace{0.1in} \mbox{(HCSW),} \label{eq4.22}
  \end{equation}
where the two amplitudes have opposite signs. Finally by using (\ref{eq2.8}), (\ref{eq4.10}), (\ref{eq4.21}) and (\ref{eq4.22}) and the estimate $\rho_c a^{3}\simeq 0.3067$ \cite{ork:fis:pan}, we obtain
  \begin{equation}
    j_{2} = 0.040 \pm 0.003, \hspace{0.1in} {\cal R}_{\mu}=-0.042 \pm 0.003 \hspace{0.1in} \mbox{(HCSW).}  \label{eq4.23}
  \end{equation}
This is, in fact, consistent with the previous, less precise estimate $-0.08\pm 0.12$ \cite{ork:fis:pan}. It should be noted that the agreement of the two, quite different procedures for estimating ${\cal R}_{\mu}$ serves as an encouraging check on the overall validity of the scaling analysis. Note that the uncertainty quoted in this estimate reflects the uncertainties only in $B$, $C^{+}$ and $\rho_c$, but not in the fitting coefficients in (\ref{eq4.22}): because of the higher order terms it is hard to provide a realistic estimate of these uncertainties. It is reasonable to believe, however, that they are likely to be no more than 20-30$\%$.

The amplitude $A_{l}$ is proportional to $l_{1}+j_{1}$: see (\ref{eq2.8}). One may then hope to estimate $l_{1}+j_{1}$, if not separately, from the fit (\ref{eq4.22}) via the same route. To do so, however, requires further information concerning the nonuniversal amplitude $D_{L}$ and the universal constant $\bar{c}_l$: see (\ref{eq2.8}) and (\ref{eq4.3}). The former can be obtained via the normalization condition $Y_{10}=1$ in the expansion (\ref{eq.Yexpand}), as for $U_{L}$; but this requires knowledge of the temperature dependence of the finite-size-scaling function $Y(x,z)$. This is also the case for estimating the universal constant $\bar{c}_l$ which contains $a_{11}$ and $a_{31}$. Hence, further investigation is needed to obtain estimates for these subdominant mixing coefficients.

\subsubsection{\label{sec4.3.2}Restricted primitive model}

The restricted primitive model (RPM) electrolyte consists of an equal number of positive and negative ions, of charges $\pm q_{0}$ and hard-core diameter $a$ interacting with each other via the Coulomb potential, $\varphi(r)=\pm q_{0}^{2}/Dr$, where $D$ is the dielectric constant of medium. For this model the Yang-Yang ratio ${\cal R}_{\mu}$ has not previously been investigated seriously. The markedly asymmetric nature of the coexistence curve hints that ${\cal R}_{\mu}$ might be large compared to the HCSW fluid. Indeed, studies of the generalized $k$-susceptibility-loci \cite{kim:fis} have suggested that ${\cal R}_{\mu}$ might be positive and nonnegligible.

To estimate ${\cal R}_{\mu}$ for the RPM (at a $\zeta = 5$ fine-lattice discretization level \cite{lui:fis:pan}) we follow the procedure used for the HCSW fluid. Thus Fig.\ \ref{fig3}(b) presents the effective susceptibility amplitude $\chi_{NN}t^{\gamma}$, along the critical isochore $\rho=\rho_c$ as obtained from grand canonical MC simulations \cite{lui:fis:pan}. An Ising fit to the precise data for the coexistence curve \cite{kim:fis:lui} can also be well achieved. Thence we find
  \begin{equation}
      B = 0.284 \pm 0.01, \hspace{0.1in}C^{+} = 0.087 \pm 0.005, \hspace{0.1in} \mbox{(RPM)}  \label{eq4.24}
  \end{equation}
with $b_\theta \simeq -1.17$ in (\ref{eq4.19a}).

As seen in Fig.\ \ref{fig2}, grand canonical MC simulations for the RPM yield strong asymmetry values. The fit of the ${\cal A}_{\mbox{\scriptsize min}}^{c}(T_{c};L)$ data to (\ref{eq4.1}) provides the amplitudes
  \begin{equation}
   j_{2}A_{j}\bar{c}_{j} \simeq 1.644, \hspace{0.1in} A_{l}\bar{c}_{l}\simeq 0.395, \hspace{0.1in} \mbox{(RPM).}  \label{eq4.25}
  \end{equation}
It is remarkable that $j_2 A_{j}$ has the opposite sign to that for the HCSW fluid while the amplitude combination is also $\sim 14$ times larger. Together with the critical density estimate, $\rho_c a^{3}\simeq 0.079$ \cite{lui:fis:pan}, we thus find
  \begin{equation}
    j_{2} = -0.35 \pm 0.07, \hspace{0.1in} {\cal R}_{\mu}=0.26\pm 0.04 \hspace{0.1in} \mbox{(RPM).}  \label{eq4.26}
  \end{equation}
In contrast to the HCSW fluid, ${\cal R}_{\mu}$ for the RPM is positive and large which seems to reflect the strongly asymmetric nature of this model electrolyte. As for the HCSW data the uncertainties in (\ref{eq4.26}) do not reflect those in (\ref{eq4.25}); however, as the combination $j_2 A_j$ is rather large and the fit in Fig.\ \ref{fig2} rather good further uncertainties of only, say 5 to $10\%$, seem likely from this source.

One should notice that the RPM studied here is, as mentioned above, a discretized version with the discretization parameter $\zeta = 5$ \cite{lui:fis:pan}. It has been shown, however, that the universal critical behavior is independent of $\zeta\,\, (\gtrsim 4)$ \cite{kim:fis3}. Furthermore, the nonuniversal critical parameters, $T_{c}$ and $\rho_{c}$, for $\zeta\geq 5$ are close to the continuum limits; the $T_c (\zeta=5)$ and $\rho_c (\zeta=5)$ are only $3\%$ and $5\%$ higher than the continuum values, respectively. Thus we believe that ${\cal R}_{\mu}(\zeta=5)$ is likely to be quite close to ${\cal R}_{\mu}(\zeta=\infty)$.

\section{\label{sec5}Scaling Algorithm for the Coexistence$\;$-curve Diameter}

Estimating the coexistence-curve diameter and, in particular, identifying its singular behavior near criticality has been a major challenge for both experiment and simulation. As mentioned in the Introduction, the presence of a nonvanishing pressure-mixing coefficient $j_{2}$ yields a $|t|^{2\beta}$ term in the diameter, that dominates the previously anticipated \cite{mer} $|t|^{1-\alpha}$ contribution. Here we present a scaling algorithm designed to enhance the estimation of the diameter near criticality, and, thereby, to improve estimates of the critical density, $\rho_c$.

The recently developed algorithm for estimating the density discontinuity, $\Delta\rho_{\infty}(T)$, utilizes the scaling relations between the $Q$-minima and their normalized locations, $y_{\mbox{\scriptsize min}}^{\pm}$: see (\ref{eq3.8}) and (\ref{eq3.12}). The asymptotically exact expression for large $L$ at fixed $T<T_c$ given in (\ref{eq20}) provides the limiting guide to construct a universal scaling function (\ref{eq3.16}) by finding optimal values for $\Delta\rho_{\infty}(T)$. The step-by-step procedure for implementing the algorithm is presented in Refs.\ \cite{kim:fis:lui} and \cite{kim:fis2}. Here we adapt the algorithm to derive the coexistence-curve diameter. For this purpose we consider the relation between ${\cal D}_{\mbox{\scriptsize min}}$ [containing $\rho_{\mbox{\scriptsize diam}}(T)$ as a variable: see (\ref{eq.diam})-(\ref{eq1.8}) and (\ref{eq22})] and $\bar{Q}_{\mbox{\scriptsize min}}$. This exhibits a rather complicated structure of finite-size corrections owing to the various mixing coefficients: see (\ref{eq3.17}); however, the relation achieves a universal form (\ref{eq22}) for large enough $L$. Thus we may hope to construct a scaling function from this limiting behavior via a scaling algorithm by optimally choosing values for $\rho_{\mbox{\scriptsize diam}}(T)$. Although the steps for the $\Delta\rho_{\infty}(T)$ algorithm are presented in detail in \cite{kim:fis2}, we recapitulate the main points in the present setting for the sake of completeness.

There are three main steps: {\bf (i)} Collect data sets of the minima of $Q_{L}$ and their locations, $\{ Q_{\mbox{\scriptsize min}}^{\pm}(T;L_{i}), \rho_{\mbox{\scriptsize min}}^{\pm}(T;L_{i}) \}$, for a range of box sizes $\{ L_{i} \}_{i=1}^{n}$ at fixed values of $T \lesssim T_c$. For this purpose, multi-histogram reweighting technique \cite{fer:swe} should be employed to generate the data at any desired temperature for a given system size. In practice, $n=3$ distinct box sizes with $L_{3} \gtrsim 1.3L_{1}$ may well suffice; to avoid the effects of the corrections to scaling one finds that one needs at least $L_{1}/a \gtrsim 8$ (where $a$ measures the size of particles).  However, this guide will surely depend on the system under consideration. {\bf (ii)} Choose a value $T=T_{0}$ sufficiently low that the two-peaked, double-Gaussian structure of ${\cal P}_{L}(T_{0};\rho)$ is well realized. In this case $\bar{Q}_{\mbox{\scriptsize min}}$ is close to zero. (For the HCSW fluid we obtained $\bar{Q}_{\mbox{\scriptsize min}}\lesssim 0.03$.) Now note that in this region the universal scaling relation between ${\cal D}_{\mbox{\scriptsize min}}$ and $\bar{Q}_{\mbox{\scriptsize min}}$ in (\ref{eq22}) is exact up to linear order. Hence, at this temperature we select a diameter estimate, say $\rho_{T_{0}}$, independent of the $L_{i}$, which leads to the best fit of 
 \begin{eqnarray}
  {\cal D}_{\mbox{\scriptsize min}}^{(i)} & \equiv & \{ \rho_{\mbox{\scriptsize m}}^{+}+\rho_{\mbox{\scriptsize m}}^{-}-2\rho_{T_0}\}/(\rho_{\mbox{\scriptsize m}}^{+}-\rho_{\mbox{\scriptsize m}}^{-})/{\cal A}_{\mbox{\scriptsize min}}\nonumber \\
  &  & \mbox{vs.} \hspace{0.2in} \bar{q}_{0}^{(i)} \equiv \bar{q}(T_{0};L_{i}) \label{eq.Dmin}
 \end{eqnarray}
to the relation (\ref{eq22}). The value $\rho_{T_{0}}$ can then be identified as an estimate for $\rho_{\mbox{\scriptsize diam}}(T_{0})$. {\bf (iii)} Increase $T_{0}$ by a small amount $\Delta T_{0}$ to $T_{1}=T_{0}+\Delta T_{0}$ and compute ${\cal D}_{\mbox{\scriptsize min}}(T_{1};L_{i})$ and $\bar{q}_{1}^{(i)} \equiv \bar{q}(T_{1};L_{i})$. In doing so, it is necessary to choose $\Delta T_{0}$ small enough that the new set $\{ \bar{q}_{1}^{(i)} \}_{i=1}^{n}$ overlaps the previous one $\{ \bar{q}_{0}^{(i)} \}_{i=1}^{n}$. Note that $\bar{q}$ increases with $T$, approaching the critical value $\bar{q}_c$ at $T_c$. When the new data set is obtained, we choose, as before, a new value, $\rho_{T_{1}}$, so that the new data set collapses optimally onto the previous one; this procedure then extends the previously computed scaling function to larger values of $\bar{q}$. Again, the new value $\rho_{T_{1}}$ can be regarded as an estimate for $\rho_{\mbox{\scriptsize diam}}(T_{1})$. These steps are repeated iteratively by increasing the temperature $T_{j}$ to $T_{j+1}=T_{j}+\Delta T_{j}$. This extends the scaling function further and generates successive estimates for $\rho_{\mbox{\scriptsize diam}}(T_{j})$ for $j=2,3,\cdots$. When criticality is approached, smaller increments $\Delta T_{j}$ are needed and high quality data prove essential. For graphical illustrations of these procedures, see Fig.\ 2 of Ref.\ \cite{kim:fis2}.

One should note that owing to the competitive nature of the singular terms in (\ref{eq3.17}), with $\kappa\simeq 0.517$ and $\lambda\simeq 0.897$, there is {\em no leading universal function} that will, in general, yield the diameter. This is in contrast to the case of the density discontinuity where there is a universal scaling function which provides an easier and faster algorithm to estimate $\Delta\rho_{\infty}(T)$: see \cite{kim:fis2}. It is therefore necessary to approach the problem case by case. In the following sections, we study the two extremal limits and compute two distinct universal scaling functions: see (\ref{eq3.18}) and (\ref{eq3.19}). First, the scaling relation (\ref{eq3.19}) derived in the absence of $j_{2}$ can be obtained from an exactly soluble decorated lattice gas model which is known to exhibit a nonvanishing $l_{1}$ but has no pressure mixing (i.e., $j_2 \equiv 0$). One may suspect that the HCSW fluid, which has been seen to exhibit a rather small pressure-mixing coefficient $j_{2}$ may be well approximated by this route. The other extremal case, in (\ref{eq3.18}), in which pressure mixing appears to strongly dominate, can be obtained by analysis of the RPM electrolyte.

\section{\label{sec6}Decorated Lattice Gas Model}

Consider a decorated lattice gas model which can be solved exactly in terms of the solution of the associated Ising model \cite{fis2,zol:mul}. The model exhibits a entropy-like singularity, $|t|^{1-\alpha}$, in the diameter but has no pressure mixing and so serves as a guide to the extremal case (\ref{eq3.19}) for the diameter.

The decorated lattice gas we study here consists of primary cells centered on sites of a basic simple cubic (sc) lattice and secondary, decorating cells centered on the bonds between the nearest neighbor pairs of sc lattice sites. All cells have equal volume, say $v_{0}$, and do not overlap. Each cell can be empty or occupied by at most one particle. Particles in nearest neighbor primary cells then interact with energy $-\epsilon$ while particles occupying neighboring primary and secondary cells interact with energy $-\lambda \epsilon$. For simplicity we will consider only $\lambda = 1$ which will suffice for our present purposes. Of course, one may consider a simpler version by only allowing the interaction between particles in nearest primary and secondary cells. However, this will not change the significant results.

The grand canonical partition function of the model can be expressed in terms of that of the ordinary lattice gas model \cite{fis2}. Let $N$ be the number of primary cells and $q$ the coordination number (with $q=6$ for the sc lattice) so that the total number of cells is $N_{\mbox{\scriptsize tot}}=(q+2)N/2$. The dimensionless activity of molecules in the decorated lattice gas is $z=v_{0}\Lambda_{T}^{-3}\exp(\mu/k_{\mbox{\scriptsize B}}T)$, where $\Lambda_{T}$ is the de Broglie wavelength and $\mu$ is the chemical potential \cite{kim4}, and we write $K\equiv \epsilon/k_{\mbox{\scriptsize B}}T$ for the coupling constant of the decorated lattice gas model. Then the grand canonical partition function can be written as
 \begin{equation}
   \Xi(z,K) = (1+z)^{qN/2}\bar{\Xi}(\bar{z},\bar{K}),
 \end{equation}
where $\bar{\Xi}$ is the corresponding partition function of the ordinary lattice gas as a function of the transformed activity and coupling constant
 \begin{eqnarray}
   \bar{z} & = & z\left(\frac{1+ze^{\lambda K}}{1+z}\right)^q, \\
   \bar{K} & = & K + \ln\left[ \frac{(1+z)(1+ze^{2\lambda K})}{(1+ze^{\lambda K})^2}\right].
 \end{eqnarray}
The critical point values, $K_{c}$ and $z_{c}$, follow by substituting $\bar{K}_{c}=4K_{c}^{\mbox{\scriptsize Is}}$ and $\bar{z}_{c}=\exp(-2qK_{c}^{\mbox{\scriptsize Is}})$ where $K_{c}^{\mbox{\scriptsize Is}}$ is the critical coupling of associated sc Ising model.

The coexistence curve of the decorated gas can be similarly obtained from that for the Ising model. The number density in the decorated gas is given by
 \begin{eqnarray}
   \rho & = & \lim_{N_{\mbox{\tiny tot}}\rightarrow\infty} N_{\mbox{\scriptsize tot}}^{-1}(\partial\ln\Xi/\partial\ln z) \nonumber \\
  & = & \frac{qz}{(q+2)(1+z)} \nonumber \\
  &  &+ \frac{2}{q+2}\left[ \bar{\rho}(\bar{z},\bar{K})\frac{\partial\ln\bar{z}}{\partial\ln z} + \bar{w}(\bar{z},\bar{K})\frac{\partial\bar{K}}{\partial\ln z}\right],  \label{eq.rho_dec}
 \end{eqnarray}
where $\bar{\rho}$ and $-\bar{w}$ are the number density and the energy per site of the ordinary lattice gas, which can be written on the phase boundary as
 \begin{equation}
   \bar{\rho}_\pm = \mbox{$\frac{1}{2}$}(1 \pm M_0 ), \hspace{0.1in} \bar{w}_\pm = \mbox{$\frac{1}{4}$}\left(\mbox{$\frac{1}{2}$}q + u \pm 6M_0 \right), \label{eq.Ising}
 \end{equation}
in which $M_0$ and $u$ are the spontaneous magnetization and the energy density of the Ising model below $T_c$; here $\pm$ represent the liquid and gas sides of the phase boundary, respectively. Note that the phase boundary in the activity-coupling plane is given by $\bar{z}(\bar{K})=\exp(-q\bar{K}/2)$. Liu and Fisher \cite{liu:fis} give approximate forms for the spontaneous magnetization and the specific heat below $T_c$ as
 \begin{eqnarray}
  M_{0}(T) & \simeq & B|t|^{\beta}(1-a_{\mbox{\scriptsize eff}}|t|^{1/2}), \nonumber \\
  C(T) & \simeq & A^{-}|t|^{-\alpha}(1-a_{\mbox{\scriptsize eff}}|t|^{1/2}) + b_{\mbox{\scriptsize eff}},  \label{eq.liu}
 \end{eqnarray}
which are valid through the whole temperature range. It is worth mentioning that one can calculate $l_{1}$ exactly, while the pressure-mixing coefficients, $j_{1}$ and $j_2$, are identically zero; but computing $l_1$ is not relevant here.

Now using (\ref{eq.rho_dec}) and (\ref{eq.Ising}), and the precise numerical results for the spontaneous magnetization, $M_0(T)$, and specific heat for the Ising model obtained via the fitting formulae (\ref{eq.liu}) together with the amplitude values in \cite{liu:fis}, we can obtain the coexistence curve, $\rho^{\pm}(T)$, for the decorated lattice gas model via (\ref{eq.rho_dec}). Thus we can construct the scaling function for $\Delta y_{\mbox{\scriptsize min}}$ presented in Fig.\ \ref{fig5} as symbols. 
\begin{figure}[ht]
\vspace{-0.95in}
\centerline{\epsfig{figure=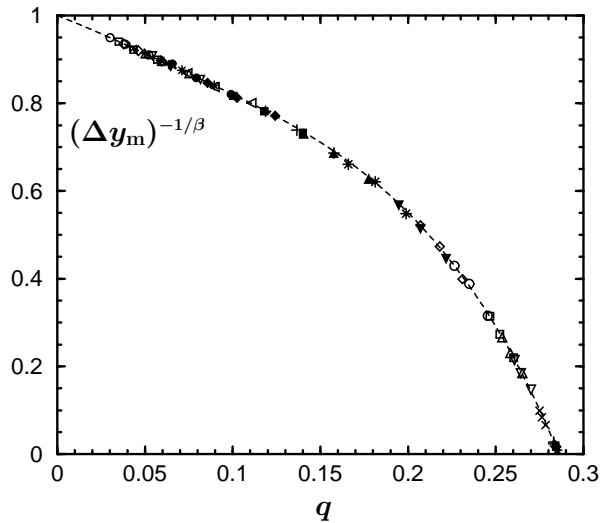,width=3.8in,angle=0}}
\vspace{-1.2in}
\caption{ Scaling plot of $(\Delta y_{\mbox{\scriptsize min}})^{-1/\beta}$ vs.\ $q\equiv \bar{Q}_{\mbox{\scriptsize min}}\ln (4/e\bar{Q}_{\mbox{\scriptsize min}})$ (with $\beta=0.326$) for the coexistence curve of the decorated lattice gas model. The various symbols depict results generated from simulations at increasing temperatures with system sizes, $L^{\ast}=8$, $9$, $10$ and $11$. For comparison, the universal scaling curve obtained previously, and represented analytically in Eq.\ (12) of \cite{kim:fis2}, is presented as a dashed curve.\label{fig5}}
\vspace{-0.1in}
\end{figure}
One should note, however, that, although the coexistence curve for the decorated gas is exactly known, building the scaling function for $\Delta y_{\mbox{\scriptsize min}}$ requires information concerning the $Q$-minima; these are finite-size quantities, which are not known exactly. Thus, to obtain the $Q$-minima and their locations, $\{ Q_{\mbox{\scriptsize min}}^{\pm}(T;L),\rho_{\mbox{\scriptsize min}}^{\pm}(T;L)\}$, for the decorated gas model, we have performed grand canonical MC simulations on lattices of dimensions $L\times L\times L$ with periodic boundary conditions. The resulting scaling curve in Fig.\ \ref{fig5} is indistinguishable from that of the HCSW fluid \cite{kim:fis2} --- see the dashed curve --- built via the recursive scaling algorithm, thus confirming our iterative scaling approach for constructing $\Delta\rho_{\infty}(T)$. We then build a scaling plot for $e_l (x)$ --- see (\ref{eq3.19}) --- by using the exact diameter of the decorated lattice gas. This curve, presented in Fig.\ \ref{fig6}, 
\begin{figure}[ht]
\vspace{-0.95in}
\centerline{\epsfig{figure=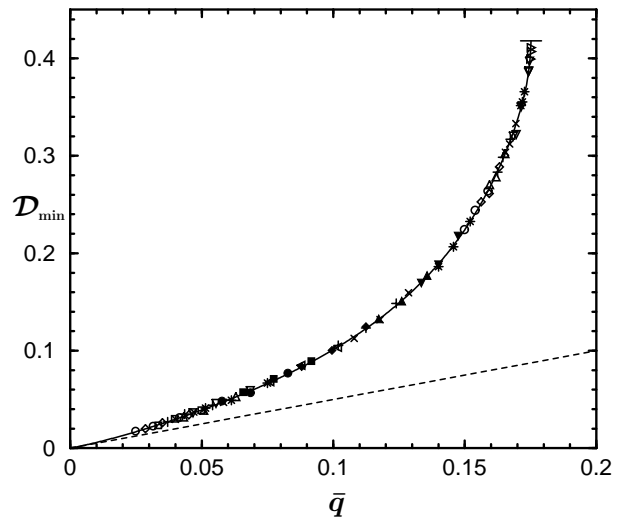,width=3.8in,angle=0}}
\vspace{-1.2in}
\caption{ Plot of the (subdominant) scaling function $e_l(x)$ vs.\ $\bar{q}\equiv \bar{Q}_{\mbox{\scriptsize min}}\ln (4/e^{2}\bar{Q}_{\mbox{\scriptsize min}}) $ for the diameter of the decorated lattice gas model with $\lambda = 1$. The dashed line represents the two-Gaussian limit (\ref{eq22}) which is asymptotically exact to linear order in $\bar{q}$, while the solid curve is a fit to (\ref{eq.Dmin_hcsw}).\label{fig6}}
\end{figure}
should represent the case when $j_{2}=0$ and $l_{1}\neq 0$, namely ${\cal D}_{\mbox{\scriptsize min}}\approx c_{l}/d_{l}$. We expect that an effective scaling curve for the HCSW fluid should be close to this one.

\section{\label{sec7}Numerical Estimates for the diameters}

Following the procedures explained we now report explicit results for the HCSW fluid and the RPM.

\subsection{\label{sec7.1}Hard-core square-well fluid}

To derive the diameter for the HCSW fluid, we start from the temperature $T_{0}^{\ast}\equiv \kb T/\epsilon = 1.10$ at which, as for the density discontinuity \cite{kim:fis2}, the double-peak Gaussian for the density distribution is quite accurate. At this temperature, we determine that $\rhodiam(T_{0})$ which leads to the best fit of ${\cal D}_{\mbox{\scriptsize min}}(T_{0};L)$ vs.\ $\bar{q}(T_{0};L)$ to the asymptotically exact two-Gaussian limit given in (\ref{eq22}): see the dashed straight line in Fig.\ \ref{fig7}(a). 
\begin{figure}[ht]
\vspace{-0.5in}
\centerline{\epsfig{figure=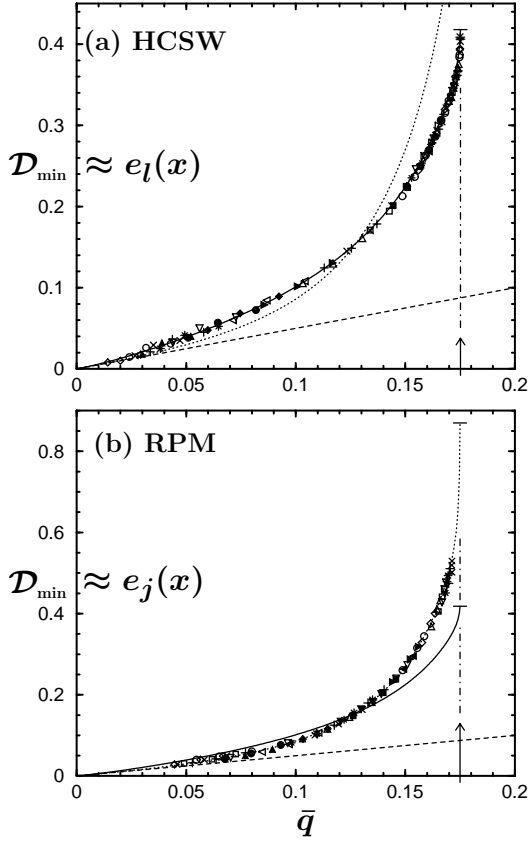,width=4.2in,angle=0}}
\vspace{-0.5in}
\caption{ Plot of scaling functions for the coexistence-curve diameters vs.\ $\bar{q}\equiv q-\bar{Q}_{\mbox{\scriptsize min}}$ for (a) the HCSW fluid and (b) the RPM. The symbols represent data at different temperatures while the dashed lines represent the exact, two-Gaussian limiting behavior (\ref{eq22}). For comparison, the plots display, as solid and dotted curves, fits to the extremal scaling curves, $e_l(x)$ and $e_j(x)$, using (\ref{eq.Dmin_hcsw}) and (\ref{eq.Dmin_rpm}). \label{fig7}}
\end{figure}
Since ${\cal D}_{\mbox{\scriptsize min}}$ depends on asymmetry, high quality data are required as stressed before. As seen in Fig.\ \ref{fig7}(a), the data exhibit relatively noisy behavior at low temperatures (possibly owing to the lack of enough histograms covering the whole range of the two-phase region at this temperature). The scatter may be compared to the analysis of the density discontinuity (see Refs.\ \cite{kim:fis:lui,kim:fis2}) where the cancelation of the uncertainties on the two sides of the coexistence region, when the average is taken, evidently results in smoother behavior. Despite the scatter in Fig.\ \ref{fig7}(a) we can estimate the diameter at $T_0$ to within about $\pm 0.2\%$ relative to $\rho_c$. Following the procedure in Sec.\ V, we then build up a scaling function numerically by increasing the temperature, $T_{j}$, and finding optimal values for $\rhodiam(T_{j})$. The function constructed this way for the HCSW fluid is presented in Fig.\ \ref{fig7}(a); it should approximate the extremal scaling limit --- see $e_l (x)$ in (\ref{eq3.19}) --- in which pressure mixing is considered negligible.

The scaling analysis of the $Q$-minima indicates that ${\cal D}_{\mbox{\scriptsize min}}(x) \approx e_l(x) = d_{l}(x)/c_{l}(x)$ in (\ref{eq3.19}) converges to a universal constant, say $C_{l}$, when $\bar{q}$ approaches the critical value $\bar{q}_c\equiv q_c-\bar{Q}_{\mbox{\scriptsize min}}^c \simeq 0.175$ as $|\bar{q}-\bar{q}_c |^{1-\alpha}$. From (\ref{eq3.10}), (\ref{eq3.15}), (\ref{eq3.19}) and (\ref{eq4.3}), this constant $C_{l}$ is given by
 \begin{equation}
  C_{l} = \bar{c}_l /d_l(0) = \bar{c}_l \left[-\frac{a_{01}}{a_{10}} + \frac{a_{20}b_{20}}{a_{10}b_{00}}\right]^{-1}, \label{eq.C_l}
 \end{equation}
where $b_{00}$ and $b_{20}$ can also be expressed in terms of the $a_{ij}$ via (\ref{eq3.5}) and (\ref{eq3.7}). Computing $C_l$, however, requires values for $a_{01}$, $a_{11}$ and $a_{31}$, which can be obtained from the temperature dependence of the finite-size scaling function $Y(x,z)$; but, at this stage these values are not known. Nevertheless, we have estimated $C_l$ via extrapolation from the HCSW data and obtained $C_l \simeq 0.418$. Using this value, we can then fit the scaling curve to the approximant
 \begin{equation}
  e_l(x) \simeq C_{l}\left[ 1-\frac{(1-q^{\prime})^{(1-\alpha)}(1+s_{1}q^{\prime}+s_{2}q^{\prime 2})}{1+t_{1}q^{\prime} + t_{2}q^{\prime 2}}\right],  \label{eq.Dmin_hcsw}
 \end{equation}
where $q^{\prime}\equiv\bar{q}/\bar{q}_{\mbox{\scriptsize c}}$ while we set $t_{1}=s_{1}-1+\alpha + \bar{q}_{\mbox{\scriptsize c}}/2C_{l}$ in order to ensure the small-$\bar{q}$ limiting behavior (\ref{eq22}). The values following from this fit are shown in both parts of Fig.\ \ref{fig7} as solid curves: the fitted coefficients are $s_1 \simeq 4.53$, $s_2 \simeq -6.61$, $s_3 \simeq 1.20$, $t_1 \simeq 3.85$, $t_2 \simeq -9.08$, and $t_3 \simeq 4.24$. Note the singular behavior at $T_c$ when $q^\prime\rightarrow 1$ implied by the $(1-q^\prime)^{1-\alpha}$ factor in (\ref{eq.Dmin_hcsw}) which yields a vertical tangent in the plot.

Figure \ref{fig8}(a) presents the diameter, $\rho_{\mbox{\scriptsize diam}}(T)$, for the HCSW fluid obtained via the scaling algorithm: see solid circles. 
\begin{figure}[ht]
\vspace{-0.5in}
\centerline{\epsfig{figure=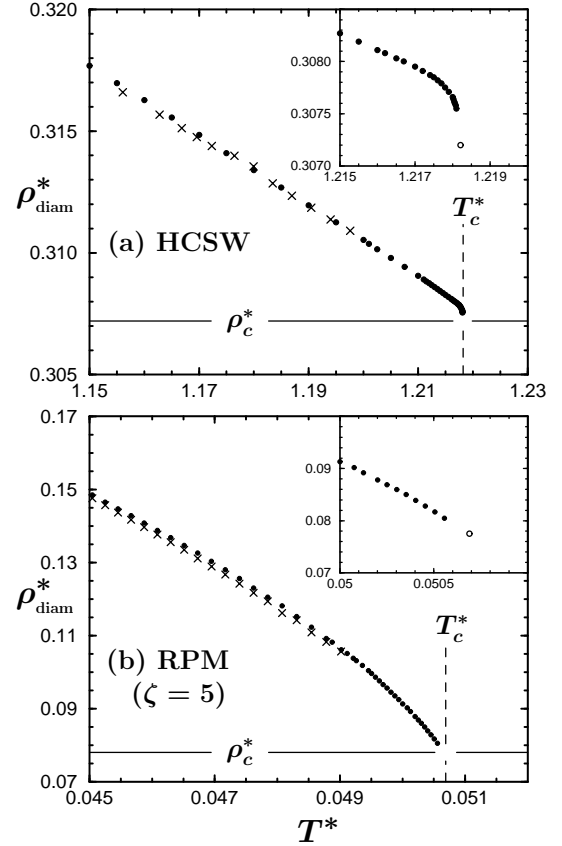,width=4.2in,angle=0}}
\vspace{-0.5in}
\caption{Plots of the coexistence curve diameters, $\rho^{\ast}_{\mbox{\scriptsize diam}}(T^\ast) \equiv\rho_{\mbox{\scriptsize diam}}a^3$ obtained via the scaling algorithm: solid circles. (a) The HCSW fluid with $T^{\ast}=\kb T/\epsilon$ and (b) the RPM (at a fine discretization level $\zeta = 5$ \cite{lui:fis:pan}) with $T^{\ast} = \kb TDa/q_{0}^{2}$. The crosses are previous estimates \cite{ork:fis:pan,lui} obtained from data for the two-peak structure of the density distribution via an equal-weight prescription. The open circles in the insets are the estimates of the critical point, $(T_c^\ast,\rho_c^\ast)$. \label{fig8}}
\end{figure}
For comparison, previous estimates obtained directly from the simulated probability distribution function of the density are presented (as crosses). In the temperature range where the equal-weight prescription works well, the estimates from the two methods agree within the uncertainties. On the other hand, the equal-weight prescription cannot provide reliable values for the diameter in the near-critical region. The current approach, however, gives $\rho_{\mbox{\scriptsize diam}}(T)$ far closer to the $T_c$. A fit to the asymptotic expansion \cite{kim:fis:ork}
 \begin{equation}
  \rho_{\mbox{\scriptsize diam}}(T) \simeq \rho_c\left[ 1 + A_{2\beta}|t|^{2\beta}+A_{1-\alpha}|t|^{1-\alpha}+A_1 t \right], \label{eq.rhod_expan}
 \end{equation}
yields the critical density as $\rho_c^\ast \equiv \rho_c a^3 = 0.3072 \pm 0.0005$ which agrees with the previous central estimate \cite{ork:fis:pan} within the uncertainties , while the amplitudes are $A_{2\beta}\simeq 0.37$, $A_{1-\alpha}\simeq  -2.14$, $A_{1}\simeq -2.52$. Note that the magnitude of $A_{2\beta}$ is almost an order smaller than that of $A_{1-\alpha}$. The sharp curvature for the diameter very close to $T_c$ reflects the $|t|^{1-\alpha}$ singularity. The magnitude seems surprisingly large but appears as an inescapable conclusion of our analysis: its validity might benefit from further investigations and, indeed, experiments.

As discussed above, we anticipate that when ${\cal D}_{\mbox{\scriptsize min}}(\bar{q})$ is constructed from data for the HCSW fluid in moderately small systems, it will match the scaling form (\ref{eq3.19}); of course, when $j_{2}$ is not identically zero, it should eventually reveal the limiting form (\ref{eq3.18}) when $L\rightarrow\infty$. In fact, this is confirmed by comparing Figs.\ \ref{fig6} and \ref{fig7}(a), where the scaling curve for the HCSW fluid (solid curve) is seen to be in good agreement with that of the decorated lattice gas model. On the other hand, for the RPM a strong $L^{-\beta/\nu}$ term arises from $j_{2}$; in this case the behavior of the ${\cal D}_{\mbox{\scriptsize min}}$ scaling function should approach $e_j(x)$ identified in the asymptotic form (\ref{eq3.18}). To check this, we move on to the calculation of the diameter for the RPM.

\subsection{\label{sec7.2}Restricted primitive model}

No changes in the previous procedures are needed to derive the diameter for the RPM. Using the data set of the $Q$-minima employed in \cite{kim:fis:lui,kim:fis2}, we construct the scaling function for ${\cal D}_{\mbox{\scriptsize min}}$ starting from the low temperature $T^\ast \equiv \kb T/(q_0^2/Da)=0.0426\simeq 0.84 T_c^\ast$. The scaling function so constructed is presented in Fig.\ \ref{fig7}(b): it clearly differs from that for the HCSW fluid. As argued, this is simply due to the relatively large pressure-mixing coefficient of the RPM. The scaling analysis yields ${\cal D}_{\mbox{\scriptsize min}}\approx e_j(x)$ in (\ref{eq3.18}) which approaches a universal constant, say $C_{j}$, as $|\bar{q}-\bar{q}_c|^{2\beta}$ when $T\rightarrow T_c -$. Just as for $C_l$, the constant $C_j$ can be written as
 \begin{equation}
  C_j = \bar{c}_j/d_j(0) = \bar{c}_j \left[-a_{10} + \frac{a_{20}b_{10}}{a_{10}b_{00}}\right]^{-1}, \label{eq.C_j}
 \end{equation}
where $b_{00}=b_0(0)$ and $b_{10}=b_1(0)$. Using (\ref{eq4.9}) and (\ref{eq4.10}) via (\ref{eq3.5}) and (\ref{eq3.6}), we have $C_{j}\simeq 0.87$. One may then fit the scaling curve to the further approximant, as for the HCSW fluid, 
 \begin{equation}
  e_j(x) \simeq C_{j}\left[ 1-\frac{(1-q^{\prime})^{2\beta}(1+s_{1}q^{\prime}+s_{2}q^{\prime 2})}{1+t_{1}q^{\prime} + t_{2}q^{\prime 2}}\right],  \label{eq.Dmin_rpm}
 \end{equation}
with $t_{1}=s_{1}-2\beta + \bar{q}_{\mbox{\scriptsize c}}/2C_{j}$. We obtain $s_1 \simeq -1.871$, $s_2 \simeq 1.050$, $s_3 \simeq -0.169$, $t_1 \simeq -2.422$, $t_2 \simeq 1.952$ and $t_3 \simeq -0.529$. The fit is presented in Fig.\ \ref{fig7} as dotted curves. As one might expect from the exponent value $2\beta$, this plot rises more sharply to the limiting constant than does $e_l(x)$ with exponent $1-\alpha$.

In Fig.\ \ref{fig8}(b) we display the present estimates (solid circles) for $\rho_{\mbox{\scriptsize diam}}(T)$ for the RPM (at a fine discretization level $\zeta=5$ \cite{lui:fis:pan}) along with the previous values (crosses) \cite{lui} estimated from the density distribution via the equal-weight prescription. The agreement is good within the uncertainties; but, as anticipated, the current approach yields reliable estimates much closer to the critical point. Extrapolation provides the critical density estimate $\rho_c^{\ast}\equiv\rho_c a^3 \simeq 0.078 (3)$ in excellent agreement with the previous estimate $\rho_c^{\ast}\simeq 0.0790 (25)$ \cite{lui:fis:pan}. Furthermore, a fit to (\ref{eq.rhod_expan}) yields $A_{2\beta}\simeq -2.03$, $A_{1-\alpha}\simeq 28.3$ and $A_1 \simeq -23.5$. It is interesting that although $2\beta\simeq 0.65$ is less than $1-\alpha\simeq 0.89$ the numerical behavior of the RPM diameter as $T$ approaches $T_c$ seems smoother than for the HCSW fluid: note, however, that the estimates for the RPM approach only to $|t| \sim 10^{-3}$ whereas those for the HCSW fluid show sharp behavior almost a decade closer to $T_c$.

\section{\label{sec8}Summary}

In summary we have provided a general method for determining the strength of the Yang-Yang anomaly from simulations of model fluids. Specifically, we have studied the isothermal minima of the fourth-order fluctuation parameter, $Q_{L}(T;\rho)$, in detail on the basis of the two-Gaussian approximation, that is exact well below $T_{c}$, and of the complete finite-size scaling theory near criticality \cite{kim:fis}. It was shown that the asymmetry factor, ${\cal A}_{\mbox{\scriptsize min}}\propto (Q_{\mbox{\scriptsize min}}^{+}-Q_{\mbox{\scriptsize min}}^{-})$, exhibits a leading term decaying as $L^{-\beta/\nu}$ and of magnitude set by the pressure-mixing coefficient, $j_{2}$, followed by a $L^{-(\Delta -1)/\nu}$ term arising from the combination of the mixing coefficients $l_{1}$ and $j_{1}$: see (\ref{eq3.11}). We then showed that precise finite-size data for ${\cal A}_{\mbox{\scriptsize min}}$ at $T_{c}$ provide a quantitative route to estimating the pressure-mixing coefficient, $j_{2}$, and thereby the Yang-Yang ratio, ${\cal R}_{\mu}$. By using universal information for the critical order-parameter distribution of $(d=3)$ Ising systems and the specific critical amplitudes of the order-parameter and the susceptibility above $T_{c}$ for the model fluids under study, one can estimate $j_{2}$ rather precisely. This method was applied to the HCSW fluid with range-to-core ratio 1.5 and the RPM electrolyte, leading to ${\cal R}_{\mu}\simeq -0.042$ and $+0.26$, respectively: see Sec.\ \ref{sec4.3}. The approach can be applied readily to any model fluid system: it will be a challenge to understand which features of a system govern the sign and magnitude of ${\cal R}_{\mu}$.

We have also presented in detail a recursive scaling algorithm using the $Q$-minima which enables one to estimate precisely the liquid-gas coexisting densities, $\rho^{\pm}(T)$, very close to $T_c$. Corresponding universal scaling functions which, in principle, can be derived from the probability distribution function, ${\cal P}_{L}(T;\rho)$, were investigated numerically via the algorithm by using grand canonical simulation data for the HCSW fluid and for the RPM \cite{ork:fis:pan,lui:fis:pan}. The two leading extremal universal scaling functions for the diameter were calculated and represented analytically in (\ref{eq.Dmin_hcsw}) and (\ref{eq.Dmin_rpm}). The algorithm yields precise results for $\rho_{\mbox{\scriptsize diam}}(T)$ in a range of temperature a decade or two closer to $T_c$ than was previously feasible: see Fig.\ \ref{fig8}. The new estimates for the critical temperatures and densities for both models agree well with the best previous estimates extrapolated from the data above $T_c$. Furthermore, the behavior of $\rho_{\mbox{\scriptsize diam}}(T)$ close to $T_c$ for the two models compares favorably with experimental data for SF$_6$ and liquid metals, respectively \cite{kim:fis4}

The new method is applicable to any model for which the order-parameter distribution can be reliably established at temperatures well below the critical temperature. To obtain successful estimates, however, one needs high quality data for the $Q$-minima and their locations.

\acknowledgements
The author is greatly indebted to Michael E.\ Fisher for his suggestions and a critical reading of this manuscript, and to  Gerassimos Orkoulas and Erik Luijten for their assistance in simulating the HCSW fluid and the RPM, respectively. The support of the National Science Foundation (through Grant No.\ CHE 03-01101) is much appreciated.


\begin{thebibliography}{32}
\bibitem[{kim({\natexlab{b}})}]{kim2}
\bibinfo{note}{Note that, owing to the choice of different scaling
  combinations, the amplitudes introduced here differ from those presented in
  Ref.\ \cite{kim:fis:lui}. It turns out that the current scaling form provides
  simpler results for the scaling behavior of $Q_{L}$, etc.}
\bibitem[{kim({\natexlab{c}})}]{kim5}
\bibinfo{note}{Based on Fig.\ \ref{fig2} of \cite{tsy:blo}, the uncertainties
  for the constants, $a$ and $c$, quoted in (\ref{eq4.7}) seem to be somewhat
  too small. If one allows somewhat larger uncertainties in these estimates,
  say about $\pm 3\%$, the implied value for $Q_{\mbox{\scriptsize min}}^{c}$
  can fall as low as $0.113$; that is only 2$\%$ larger than the estimated
  value $0.1107$ for the HCSW fluid.}
\bibitem[{kim({\natexlab{d}})}]{kim6}
\bibinfo{note}{Note that these values are obtained by linearly extrapolating
  the estimates of \cite{liu:fis} to the currently preferred exponent values
  presented in Table \ref{tab1}.}
\bibitem[{kim({\natexlab{e}})}]{kim4}
\bibinfo{note}{One should not confuse the symbol $z$ here with the scaling
  variable defined in (26); in this section, $z$ is used only to represent the
  fugacity.}
\end{thebibliography}
\bibliographystyle{apsrev}

\end{document}